\documentclass[twocolumn]{aastex62}
\graphicspath{{./}{figures/}}
\usepackage{xcolor}
\usepackage{graphicx}
\usepackage{soul}
\shorttitle{GC pulsar search with ALMA}
\shortauthors{Liu et al.}
\usepackage[encapsulated]{CJK}
\newcommand{\cntext}[1]{\begin{CJK}{UTF8}{gbsn}#1\end{CJK}}
\usepackage{lineno}

\begin{document}

\author[0000-0002-2953-7376]{Kuo Liu}
\affiliation{Max-Planck-Institut f\"ur Radioastronomie, Auf dem H\"ugel 69, D-53121 Bonn, Germany}

\author[0000-0003-3922-4055]{Gregory Desvignes}
\affiliation{Max-Planck-Institut f\"ur Radioastronomie, Auf dem H\"ugel 69, D-53121 Bonn, Germany}
\affiliation{LESIA, Observatoire de Paris, Universit\'e PSL, CNRS, Sorbonne Universit\'e, Universit\'e de Paris, 5 place Jules Janssen, 92195 Meudon, France}

\author[0000-0001-6196-4135]{Ralph P. Eatough}
\affiliation{National Astronomical Observatories, Chinese Academy of Sciences, 20A Datun Road, Chaoyang District, Beijing 100101, P. R. China}

\author[0000-0002-5307-2919]{Ramesh Karuppusamy}
\affiliation{Max-Planck-Institut f\"ur Radioastronomie, Auf dem H\"ugel 69, D-53121 Bonn, Germany}

\author[0000-0002-4175-2271]{Michael Kramer}
\affiliation{Max-Planck-Institut f\"ur Radioastronomie, Auf dem H\"ugel 69, D-53121 Bonn, Germany}

\author[0000-0001-8700-6058]{Pablo Torne}
\affiliation{Institut de Radioastronomie Millim\'etrique (IRAM), Avda. Divina Pastora 7, Local 20, 18012 Granada, Spain}
\affiliation{Max-Planck-Institut f\"ur Radioastronomie, Auf dem H\"ugel 69, D-53121 Bonn, Germany}

\author[0000-0002-7416-5209]{Robert Wharton}
\affiliation{Max-Planck-Institut f\"ur Radioastronomie, Auf dem H\"ugel 69, D-53121 Bonn, Germany}

\author[0000-0002-2878-1502]{Shami Chatterjee}
\affiliation{Cornell Center for Astrophysics and Planetary Science, Cornell University, Ithaca, NY 14853, USA}

\author[0000-0002-6156-5617]{James M. Cordes}
\affiliation{Cornell Center for Astrophysics and Planetary Science, Cornell University, Ithaca, NY 14853, USA}

\author[0000-0002-2079-3189]{Geoffrey B. Crew}
\affiliation{Massachusetts Institute of Technology Haystack Observatory, 99 Millstone Road, Westford, MA 01886, USA}

\author{Ciriaco Goddi}
\affiliation{Department of Astrophysics, Institute for Mathematics, Astrophysics and Particle Physics (IMAPP), Radboud University, P.O. Box 9010, 6500 GL Nijmegen, The Netherlands}
\affiliation{Leiden Observatory - Allegro, Leiden University, P.O. Box 9513, 2300 RA Leiden, The Netherlands}

\author[0000-0001-5799-9714]{Scott M. Ransom}
\affiliation{National Radio Astronomy Observatory, Charlottesville, VA 22903, USA}

\author{Helge Rottmann}
\affiliation{Max-Planck-Institut f\"ur Radioastronomie, Auf dem H\"ugel 69, D-53121 Bonn, Germany}

\author[0000-0002-9791-7661]{Federico Abbate}
\affiliation{Max-Planck-Institut f\"ur Radioastronomie, Auf dem H\"ugel 69, D-53121 Bonn, Germany}

\author[0000-0003-4056-9982]{Geoffrey C. Bower}
\affiliation{Institute of Astronomy and Astrophysics, Academia Sinica, 645 N. A'ohoku Place, Hilo, HI 96720, USA}

\author[0000-0002-2322-0749]{Christiaan D. Brinkerink}
\affiliation{Department of Astrophysics, Institute for Mathematics, Astrophysics and Particle Physics (IMAPP), Radboud University, P.O. Box 9010, 6500 GL Nijmegen, The Netherlands}

\author[0000-0002-2526-6724]{Heino Falcke}
\affiliation{Department of Astrophysics, Institute for Mathematics, Astrophysics and Particle Physics (IMAPP), Radboud University, P.O. Box 9010, 6500 GL Nijmegen, The Netherlands}
\affiliation{Max-Planck-Institut f\"ur Radioastronomie, Auf dem H\"ugel 69, D-53121 Bonn, Germany}

\author[0000-0002-4151-3860]{Aristeidis Noutsos}
\affiliation{Max-Planck-Institut f\"ur Radioastronomie, Auf dem H\"ugel 69, D-53121 Bonn, Germany}

\author[0000-0001-7520-4305]{Antonio Hern\'andez-G\'omez}
\affiliation{Max-Planck-Institut f\"ur Radioastronomie, Auf dem H\"ugel 69, D-53121 Bonn, Germany}

\author[0000-0001-7369-3539]{Wu Jiang  (\cntext{江悟})}
\affiliation{Shanghai Astronomical Observatory, Chinese Academy of Sciences, 80 Nandan Road, Shanghai 200030, China}
\affiliation{Key Laboratory of Radio Astronomy, Chinese Academy of Sciences, Nanjing 210008, China}

\author[0000-0002-4120-3029]{Michael D. Johnson}
\affiliation{Center for Astrophysics $\vert$ Harvard \& Smithsonian, 60 Garden Street, Cambridge, MA 02138, USA}
\affiliation{Black Hole Initiative at Harvard University, 20 Garden Street, Cambridge, MA 02138, USA}

\author[0000-0002-7692-7967]{Ru-Sen Lu (\cntext{路如森})}
\affiliation{Shanghai Astronomical Observatory, Chinese Academy of Sciences, 80 Nandan Road, Shanghai 200030, China}
\affiliation{Kavli Institute for Astronomy and Astrophysics, Peking University, Beijing 100871, China}
\affiliation{Max-Planck-Institut f\"ur Radioastronomie, Auf dem H\"ugel 69, D-53121 Bonn, Germany}

\author[0000-0002-3523-9156]{Yurii Pidopryhora}
\affiliation{Max-Planck-Institut f\"ur Radioastronomie, Auf dem H\"ugel 69, D-53121 Bonn, Germany}

\author[0000-0002-1330-7103]{Luciano Rezzolla}
\affiliation{Institut f\"ur Theoretische Physik, Max-von-Laue-Strasse 1, 60438 Frankfurt, Germany}
\affiliation{Frankfurt Institute for Advanced Studies, Ruth-Moufang-Strasse 1, 60438 Frankfurt, Germany}
\affiliation{School of Mathematics, Trinity College, Dublin 2, Ireland}

\author[0000-0002-1334-8853]{Lijing Shao}
\affiliation{Kavli Institute for Astronomy and Astrophysics, Peking University, Beijing 100871, China}
\affiliation{Max-Planck-Institut f\"ur Radioastronomie, Auf dem H\"ugel 69, D-53121 Bonn, Germany}
\affiliation{National Astronomical Observatories, Chinese Academy of Sciences, Beijing 100012, China}

\author[0000-0003-3540-8746]{Zhiqiang Shen (\cntext{沈志强})}
\affiliation{Shanghai Astronomical Observatory, Chinese Academy of Sciences, 80 Nandan Road, Shanghai 200030, China}
\affiliation{Key Laboratory of Radio Astronomy, Chinese Academy of Sciences, Nanjing 210008, China}

\author[0000-0003-4058-2837]{Norbert Wex}
\affiliation{Max-Planck-Institut f\"ur Radioastronomie, Auf dem H\"ugel 69, D-53121 Bonn, Germany}

\title{An 86-GHz search for Pulsars in the Galactic Center with the Atacama Large Millimeter/submillimeter Array}

\correspondingauthor{K.~Liu}
\email{kliu@mpifr-bonn.mpg.de}

\begin{abstract}
We report on the first pulsar and transient survey of the Galactic Center (GC) with the Atacama Large Millimeter/submillimeter Array (ALMA). The observations were conducted during the Global Millimeter VLBI Array campaign in 2017 and 2018. We carry out searches using timeseries of both total intensity and other polarization components in the form of Stokes parameters. We incorporate acceleration and its derivative in the pulsar search, and also search in segments of the entire observation to compensate for potential orbital motion of the pulsar. While no new pulsar is found, our observations yield the polarization profile of the GC magnetar PSR~J1745$-$2900 at mm-wavelength for the first time, which turns out to be nearly 100\% linearly polarized. Additionally, we estimate the survey sensitivity placed by both system and red noise, and evaluate its capability of finding pulsars in orbital motion with either Sgr~A* or a binary companion. We show that the survey is  sensitive to only the most luminous pulsars in the known population, and future observations with ALMA in Band-1 will deliver significantly deeper survey sensitivity on the GC pulsar population. 
\end{abstract}


\section{Introduction} \label{sec:intro}
The inner parsec of our Galaxy has been an exciting laboratory in particular for probing gravity in the past decades. This is mainly thanks to the Nobel-Prize-winning discovery of the supermassive black hole at the Galactic Center \citep{eg96,gkm+98}, whose corresponding radio source is known as Sgr~A* \citep{lbk+93,kzw+93,fma00,pge+15}. Infrared observations have revealed that Sgr~A* is surrounded by a dense group of orbiting young and old stars, commonly referred to as the S-star cluster \citep{egh+93,sog+02,gbd+03,ega+05}. Monitoring of the S-stars has led to a precise determination of the mass of Sgr~A* \citep{eg96,gkm+98,bgs+16,gpe+17}, and recent measurements of gravitational redshift and Schwarzschild precession of the S2 star orbit \citep{aaa+18,dhg+19,gaa+20}. Near-infrared and X-ray observations of flares occurring around Sgr~A* imply a significant rotation of the supermassive black hole \citep{gso+03,agpp04}.

In parallel, radio imaging studies using very long baseline interferometry at millimeter wavelengths have demonstrated the compactness of Sgr~A*, yielding an estimate on the intrinsic angular size of $40$--$60$\,$\mu$as which corresponds to only $4$--$6$ Schwarzschild radii \citep{dwr+08,fdb+11,lkr+18,jnp+18}. These crude estimates are expected to soon be replaced by 1.3-mm images of Sgr~A* with observations of the Event Horizon Telescope \citep{eaa+19}. 

Pulsars in orbit around Sgr~A*, once found, will allow 
follow-up pulsar timing observations that can  yield estimates for the mass, spin and possibly the quadrupole moment of Sgr~A* with unprecedented precision \citep{wk99,kbc+04,lwk+12,pwk16,le17,dyp+18}. The assembly of all aforementioned experiments, will deliver a multi-messenger experiment on gravity \citep{pwk16,gfk+17}, testing the fundamental principles in black-hole physics of general relativity, i.e., the cosmic censorship conjecture and no-hair theorem. Finding pulsars in the inner parsec will in addition provide a direct probe into the magneto-ionized environment in the vicinity of Sgr~A*, and shed light on the structure of its accretion flows \citep{efk+13,dep+18}.

Starting in the 1990s, pulsar searches toward the GC have so far discovered six radio-emitting neutron stars within half a degree from Sgr~A* \citep{jkl+06,dcl09,efk+13}. The GC magnetar  PSR~J1745$-$2900, initially identified in X-ray \citep{mgz+13,rep+13}, is the only one out of the six inhabiting the innermost parsec. The number of discoveries is well below the substantial expected population of pulsars in the GC region \citep[e.g.,][]{pl04,wcc+12,cl14,zly14}. 

The main challenge of the searches at decimeter wavelengths, as well summarized in \cite{cl97}, \cite{mkf+10}, \cite{jkl+06} and Eatough et al. (2021 submitted), is the high brightness temperature of the GC region \citep{rfr+90} and the temporal scattering of the electromagnetic pulses when they propagate from the GC through the dense ionized interstellar medium (ISM) to the Earth. Though the GC magnetar PSR~J1745$-$2900 exhibits much less temporal scattering than previously predicted \citep{sle+14} and similar angular broadening as Sgr~A* \citep{bdd+14,bdd+15}, it is still debatable whether one or a few lines of sight are representative given the spatial complexity of scattering materials in the GC \citep{lc98,elc+15}. The combined angular and temporal broadening of PSR~J1745$-$2900 suggest that the angular broadening is dominated by plasma far from Sgr~A* \citep{bdd+14}. The scattering of the other GC pulsars also suggests a complex picture, with the material arising at multiple long distances from the GC \citep{ddb+17}. Imaging observation of the extragalactic background source, G359.087+018, additionally supports the existence of patchy scattering structure in the GC \citep{lag+99}. Thus, strong scattering could still heavily affect the detectability of GC pulsars in low-frequency searches \citep{sle+14,mk15}, motivating searches at shorter wavelengths where temporal scattering ($\propto \lambda^4$) and the background noise of the GC is much less significant. 

Very recently, Eatough et al. (2021 submitted) and \cite{tde+21} have reported pulsar search efforts at observing wavelengths ranging from 6\,cm down to 2\,mm. At these wavelengths, sensitivity becomes a more important limiting factor given that pulsars are typically steep-spectrum radio sources. Moreover, significant power baseline variation is commonly seen in observations at these wavelengths, as a result of fluctuation in troposphere contribution to the overall system temperature \citep[e.g.,][]{lyw+19}. This effect is known to degrade de facto survey sensitivity, especially for long-period pulsars ($P\gtrsim0.1s$) \citep{lbh+15}.

Therefore, for high-frequency surveys it is crucial to have highly sensitive facilities with relatively steady tropospheric weather conditions. In light of such, the Atacama Large Millimeter/submillimeter Array (ALMA) is an ideal instrument for pulsar searching in the GC. While normally being operated as an interferometer, ALMA can now also coherently combine individual antennas and form a tied-array beam towards a single pointing, which was developed under the ALMA Phasing Project \citep[APP;][]{mcd+18}. The ALMA phasing system currently allows a coherent addition of up to 50 (usually 41) 12-m antennas, which delivers a sensitivity equivalent to maximally a 84-m single dish at mm-wavelengths. The excellent site location and the feature of being an interferometer, minimize the level of power baseline variation and guarantee the least possible radio interference (see e.g., the ALMA Technical Handbook\footnote{\url{https://almascience.eso.org/documents-and-tools/cycle8/alma-technical-handbook}}). Additionally, since pulsars are typically highly polarized radio sources, the assurance of polarization purity in the ALMA observations \citep{gmm+19}, provides a complementary probe for the surveys as will be discussed below. Recently, a dedicated pulsar observing mode with the phased ALMA has been developed under the {\em ALMA Pulsar Mode Project} (APMP)\footnote{\url{http://hosting.astro.cornell.edu/research/almapsr/.}}, which demonstrates the strength and the quality of data with ALMA for pulsar studies \citep{lyw+19}.

In this paper, we report the first pulsar and transient search of the GC with ALMA. The paper is organized as follows. Section~\ref{sec:obs} describes details of the observations carried out. Results of the searches conducted can be found in Section~\ref{sec:result}. We further discuss the sensitivity of the survey in Section~\ref{sec:sensi} and its effectiveness in probing the GC pulsar population in Section~\ref{sec:dis}. A brief conclusion is presented in Section~\ref{sec:conc}.

\section{Observations and data processing} \label{sec:obs}
\begin{figure}
\centering
\includegraphics[scale=0.6]{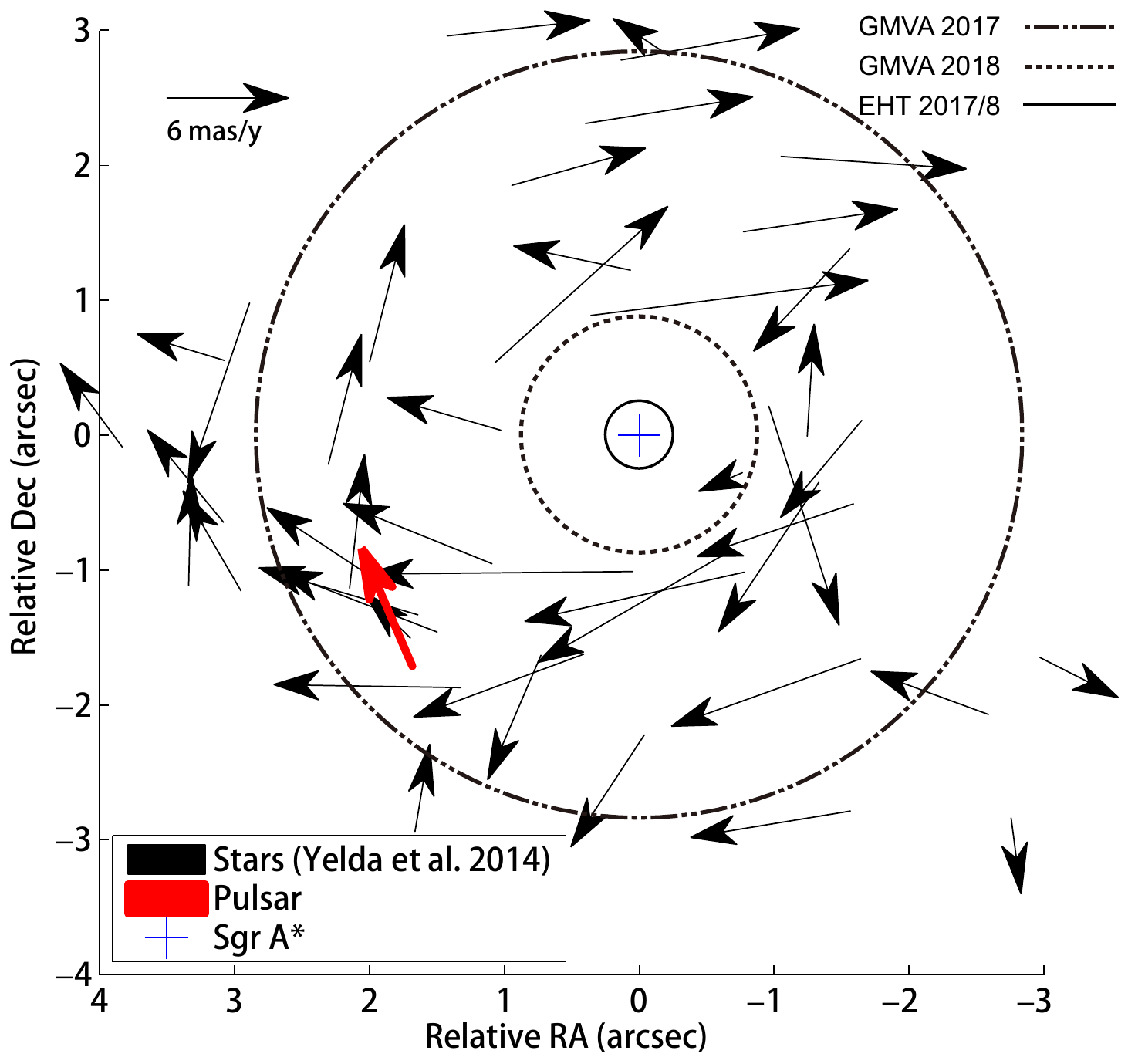}
\caption{ALMA synthetic beams (full width at half maximum) in the GMVA 2017 and 2018 campaigns (at 86\,GHz), along with that from the Event Horizon Telescope (EHT) 2017 and 2018 observations (at 228\,GHz) as a comparison \citep{eaa+19}. The black arrows represent the position and the direction of motion of the S-stars and the GC magnetar PSR~J1745$-$2900 is marked by the red arrow. The size of the ALMA synthetic beam (in arcsec,  $\mathrm{''}$) is given by $\theta=76\mathrm{''}/(L_{\rm max}f)$ as given in ALMA Technical Handbook, where $L_{\rm max}$ is the maximum baseline (in km) and $f$ is the observing frequency (in GHz). For GMVA 2017, 2018 and EHT 2017/8 observations, $L_{\rm max}$ is 0.155, 0.5, and 0.46\,km, respectively, which in turn gives $\theta\simeq5.7\mathrm{''}$, 1.8$\mathrm{''}$ and 0.72$\mathrm{''}$. This figure is based on a reproduction of Figure~3 in \cite{bdd+15}. \label{fig:ALMAbeam}}
\end{figure}

Observations of Sgr~A* ($\alpha_\mathrm{J2000} = 17^\mathrm{h}45^\mathrm{m}40^\mathrm{s}.0361$, $\delta_\mathrm{J2000} = -29^\mathrm{\circ}00\mathrm{'}28\mathrm{''}.168$) with ALMA were carried out at 3.5\,mm during the Global Millimeter VLBI Array (GMVA) campaign on April 03, 2017, April 14 and 17, 2018 \citep[][Issaoun et al. 2021, submitted]{ijb+19}. The array was coherently combined using the phasing mode developed by the APP. The sky coverage of the resulting synthetic beams in these two campaigns is shown in Figure~\ref{fig:ALMAbeam}. At each epoch, the duration of the track was approximately 6 hours in total, and divided into individual scans of approximately 3$-$6\,min each, switching between Sgr~A* and calibrator sources that include 3C\,279, NRAO\,530, and J1924$-$2914 (OV\,$-$236). Baseband voltage data of 2-GHz bandwidth centered at 86.268\,GHz were recorded in 2-bit samples on spinning discs in MARK\,6 recorders with dual polarization. We refer to \cite{gmm+19} for more details on phasing observations and data acquisition of ALMA.

The data on Sgr~A* were later processed into intensity timeseries and written in PSRFITS format in search mode using the software tool \texttt{vdif2psrfits} developed under the APMP \citep{lyw+19}. The PSRFITS product contains timeseries of four Stokes parameters stored in 32-bit float samples, with time and frequency resolution of 8\,$\mu$s and 62.5\,MHz, respectively. Though the converted PSRFITS data did not proceed through the Level 2 Quality Assurance (QA2) stage for gain and leakage calibration, both are
no more than a few percent for the observations \citep[][; Goddi et al. 2020, submitted]{gmm+19}. This means that the polarization of the data is trustworthy at a similar level of precision.

During the post-processing, we noticed a power drop-off feature in the data, with a period of 18.192\,s that coincides with the phasing cycle of the array \citep[see][for details]{gmm+19}. We accordingly developed a dedicated scheme to mitigate
this effect, 
which has significantly improved the data quality in terms of searching for time-domain signals. More details can be found in Appendix~\ref{app:sys}. It was also found that the data recorded on April 14, 2018 were heavily affected by packet losses (up to 20\% per scan for one polarization, in a number of small chunks) during the data recording. Thus, we focused on data from April 03, 2017 and April 17, 2018 for the analysis in the rest of the paper.

\section{Results} \label{sec:result} 
\subsection{Periodicity search} \label{ssec:psearch}
We searched for periodic signals in the dataset using the Fourier domain technique incorporated in the \textsc{presto} software package \citep{rem02,ar18}. 
For each epoch, the timeseries from individual scans on the Sgr~A* were coherently connected (with samples equal to the mean filled in the gaps), re-sampled to a time resolution of 32\,$\mu$s, and de-dispersed with a dispersion measure (DM) of 1,700\,cm$^{-3}$\,pc using \textsc{presto}'s \texttt{prepdata} program. The de-dispersion corrects for the frequency-dependent time delay caused by the ISM along the line of sight, which, following \cite{lk05}, can be calculated as
\begin{equation}
    \delta\tau=4.15~{\rm ms} \times(f^{-2}_{\rm low}-f^{-2}_{\rm high})\times{\rm DM}
\end{equation}
where $f_{\rm low}$, $f_{\rm high}$ are the low and high ends of the frequency band in units of GHz and $\delta\tau$ is in ms. The DM value was chosen based on the measurement of the GC magnetar PSR~J1745$-$2900 which gave ${\rm DM}\simeq1,770$\,cm$^{-3}\,$pc \citep{efk+13}. The smearing time across the entire 2-GHz band centered at 86\,GHz would be approximately 46\,$\mu$s, slightly higher than our time resolution, if the dispersion delay were not compensated for. Subtracting a DM of 1,700\,cm$^{-3}$\,pc should therefore largely mitigate the DM smearing and leave the remaining part well below the sample interval for GC pulsars with  DMs similar to that of PSR~J1745$-$2900.

\begin{figure}
\centering
\includegraphics[scale=0.75]{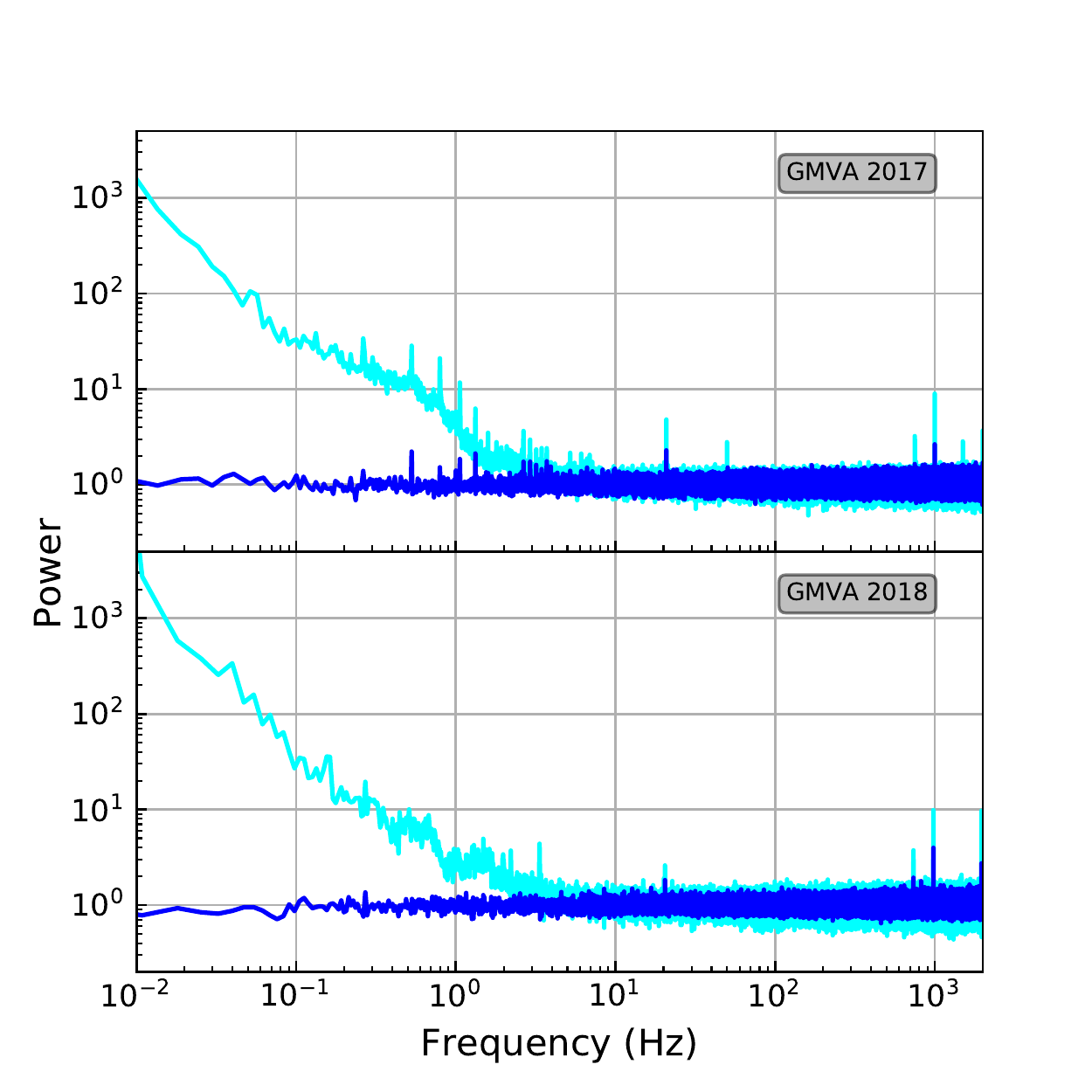}
\caption{Fourier power spectrum of the total intensity dataset from GMVA 2017 and 2018 campaigns, respectively, before (cyan) and after (blue) the application of spectrum whitening. \label{fig:pspec}}
\end{figure}

Power baseline variations are commonly seen in mm-wavelength observations as a consequence of system temperature fluctuations on timescales down to the order of seconds \citep{lyw+19}. This results in a reddening of the Fourier power spectrum and is already a known limiting factor to sensitivity in pulsar searches at longer wavelengths \citep[e.g.,][]{lbh+15}. To mitigate this impact in our dataset, we used \textsc{presto}'s \texttt{rednoise} program to subtract a running median in the Fourier power spectrum before carrying out the search. The block size in Fourier frequency bins starts from 6 bins and is increased to 100 bins at a frequency of 10\,Hz after which it remains constant. This has effectively whitened the Fourier power spectrum as shown in Figure~\ref{fig:pspec}, which  subsequently improves the sensitivity of the search to slow pulsars (e.g., $P\gtrsim0.1$\,s).

We then used \textsc{presto}'s \texttt{accelsearch} program to search for periodicities in the data. 
To compensate for potential orbital motion of a pulsar, in addition to the spin period we also varied the first and second period derivatives (acceleration and ``jerk'') to maximize sensitivity \citep[e.g.,][]{rem02,blw13,ekl+13}. 
Here we considered two possible scenarios, a pulsar in orbit with Sgr~A* or in a close binary system with another stellar-mass companion. For the former we used \texttt{accelsearch -zmax 1200 -wmax 40} and searched the full length of the dataset, while for the latter we subdivided the dataset uniformly into three segments and used \texttt{accelsearch -zmax 1200 -wmax 1500} for each of the search individually. The segmented search provides additional sensitivity to pulsars in a compact orbit with a companion or in a close orbit with Sgr~A* \citep[e.g.,][]{jk91}. More details of the search sensitivity in terms of orbital motion for the applied options in \texttt{accelsearch} can be found in Section~\ref{ssec:bnry}. We repeated the searches described above four times, using timeseries of the four Stokes parameters, respectively. The searches carried out in $Q$, $U$, $V$ (but not in $L=\sqrt{Q^2+U^2}$ as discussed in Appendix~\ref{app:sto}) are complementary to that in total intensity $I$, since the variation of the power baseline is not supposed to be polarized. We also corrected for the rotation of the linear feeds in parallactic angle, which otherwise would cause sinusoidal variations in the $Q$ and $U$ components. The Faraday rotation across the band, if adopting the rotation measure of PSR~J1745$-$2900, is calculated to be approximately 4.2\,deg. This suggests that the existence of rotation measure is likely to only have a minimal impact on the search sensitivity in $Q$ and $U$ (though other pulsars in the neighborhood may well have quite different RMs than the magnetar). The detection significance (S/N) of the Fourier power for a pulsar candidate was calculated by summing a number of 16 Fourier harmonics. For each candidate with $\rm S/N>2$, we folded the total intensity timeseries with respect to its characteristics (period, period derivative, and jerk) and visually inspected the constructed signal.

\begin{figure*}
\hspace*{-0.9cm}
\includegraphics[scale=0.42]{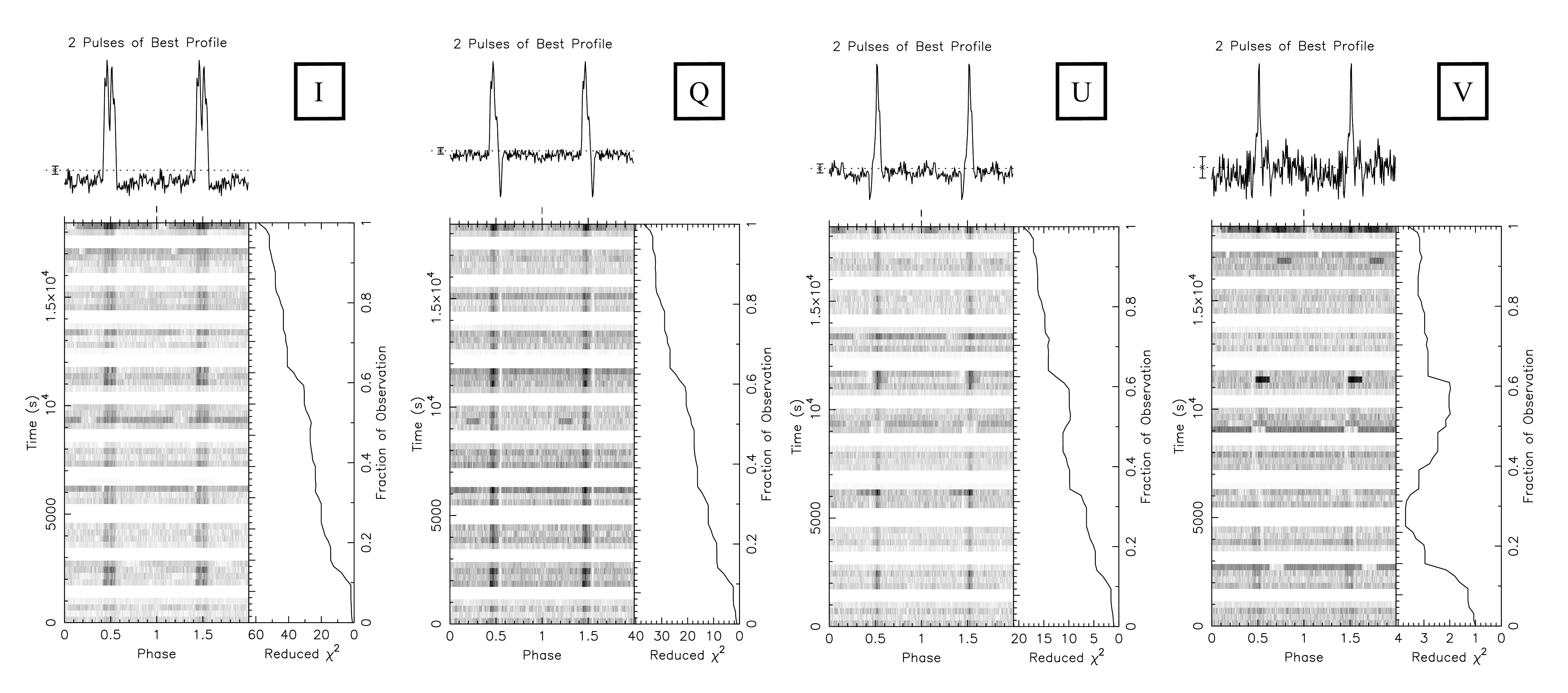}
\caption{Detection of the GC magnetar PSR~J1745$-$2900 from the blind search with the full length of the 2017 campaign using the four Stokes parameters. In all of the four cases, the magnetar was found as the candidate with the highest detection significance from the search. \label{fig:magpresto2017}}
\end{figure*}

Table~\ref{tab:Ncand} summarizes the number of candidates with S/N$>2$ from the search. It can be seen that overall the segmented search produces significantly more candidates, mainly due to the additional parameter space explored. The searches in Stokes $Q$, $U$, and $V$ yield substantially fewer  candidates, reflecting the smaller contamination of their (zero-mean) timeseries by baseline variations. The GC magnetar J1745$-$2900 was blindly detected in all searches conducted in the 2017 campaign, except for the segmented searches in $V$ due to a sensitivity limit. Figure~\ref{fig:magpresto2017} shows the detection of J1745$-$2900 in all four Stokes parameters from the searches using the full length of the observation. The S/Ns are in turn 28, 33, 22, and 4 for $I$, $Q$, $U$, and $V$, which demonstrates the success of finding pulsars with polarization components. The GC magnetar was not found in the blind search from the 2018 campaign, largely due to the fact that it was not covered by the 
synthesized beam of ALMA during the observation as illustrated in Figure~\ref{fig:ALMAbeam}. Folding the data with a pulsar ephemeris obtained from low-frequency monitoring observations (Desvignes et al.\ in prep.) revealed a weak and time-varying detection, indicating that its signal was collected in the side-lobe of the array which rotated with respect to the sky along the phase center during the tracking. The average strength of the signal is roughly an order of magnitude lower than that from the 2017 observation, so a non-detection from the blind search is expected as it falls under the search sensitivity as will be discussed in Section~\ref{ssec:lim}. Apart from the GC magnetar, no other evident pulsar candidate has been found. 

\begin{table}[]
    \centering
    \begin{tabular}{ccc}
    \hline
    \hline
Search &  $N_{\rm I}$ & $N_{\rm Q,U,V}$\\
    \hline 
2017, full & 153  & 61 \\
2017, segments & 2165 & 111 \\
\hline
2018, full & 217  & 84 \\
2018, segments & 2449 & 1693 \\
\hline
    \end{tabular}
    \caption{Number of candidates obtained from both the full-length and the segmented searches, in timeseries of total intensity and the other three Stokes components in total, respectively. }
    \label{tab:Ncand}
\end{table}

\subsection{Single-pulse search} \label{ssec:sglsearch}
In addition to periodic signals, we have also searched for 
single bright pulses from pulsars or from transient sources such as those that emit fast radio bursts (FRBs). 
Note that for searches at decimeter wavelengths, the dispersion sweep and multi-beam comparisons are crucially useful criteria for distinguishing terrestrial signals from possible astrophysical events \citep[e.g.,][]{bbe+11};
however these tools are not available with our mm observations. In light of such constraints, we developed a search scheme using polarization property as the main criterion. As seen in previous observations, bright individual pulses 
commonly exhibit significant polarization at decimeter wavelengths \citep{lk05,pbj+16}. This has also been indicated in a few pulsars at 3-mm wavelength \citep{tde+17,lyw+19}. Thus, coincidence of detection in both total intensity and one (or more) of the other polarization components can be applied during the candidate selection to find actual astronomical signals, assuming that terrestrial signals caught during the observations were not significantly polarized. Therefore, we carried out single-pulse search to timeseries in total intensity $I$, linear $L$ and circular $V$ polarization, respectively, with the \textsc{single\_pulse\_search.py} program in \textsc{presto}. Since the intrinsic $V$ component of a pulsar can be of either positive or negative sign, for the search in $V$ we first subtracted the mean of the timeseries and carried out an additional search to the data multiplied by $-1$. Then we checked the time of event for all candidates above 7-$\sigma$ significance, and kept those which were detected simultaneously in $I$ and either $L$ or $V$. In total, we have detected three events (all from $I$ and $L$ component) from the 2017 campaign and none from 2018. All event times coincide with the phase of the pulse profile of PSR~J1745$-$2900, suggesting that these events are very likely to be individual pulses of the GC magnetar (see Section~\ref{ssec:mag}).

\subsection{The magnetar signal} \label{ssec:mag}
\begin{figure}
\centering
\includegraphics[scale=0.6]{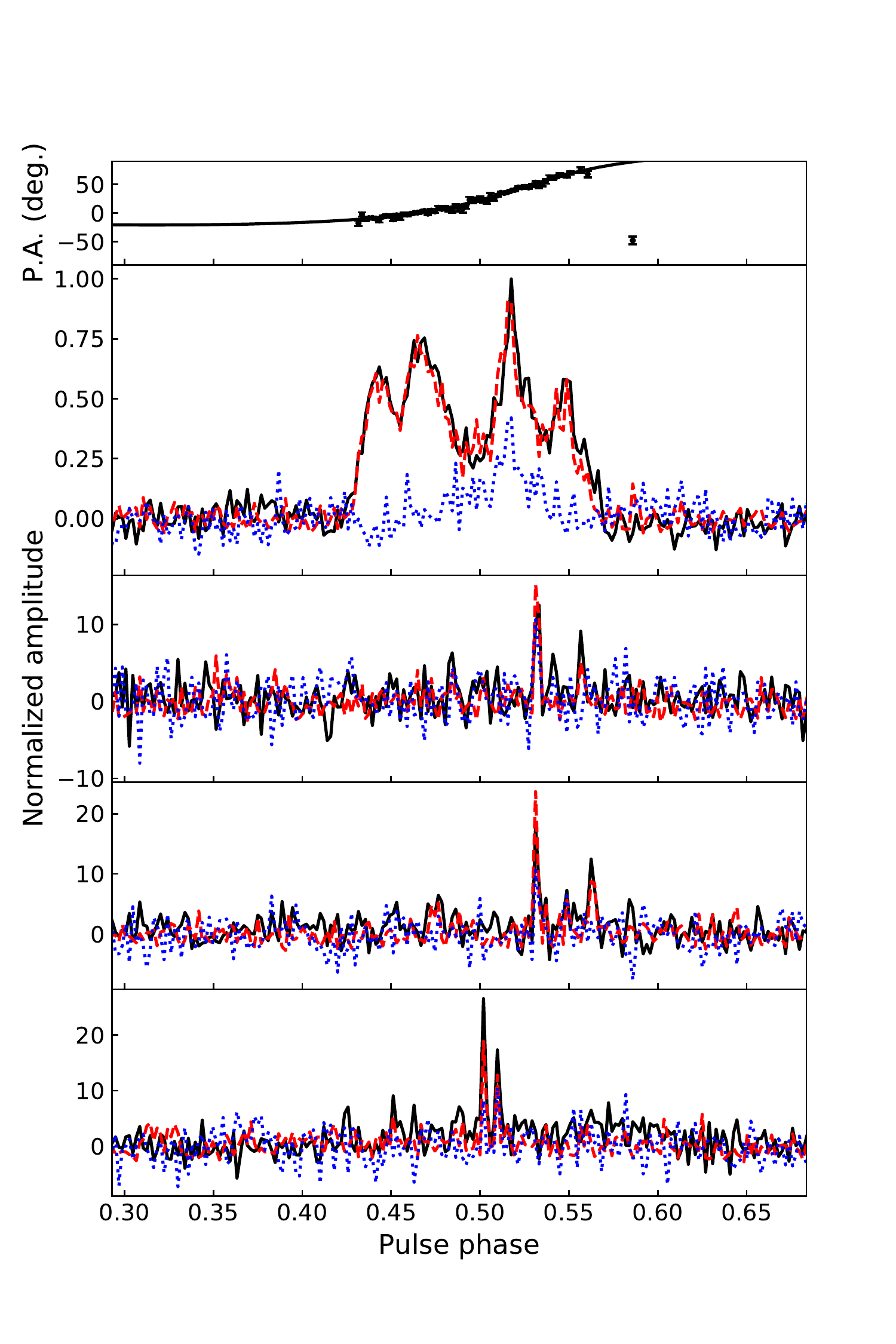}
\caption{Polarimetric signals of the GC magnetar from the 2017 campaign. The top two panels are the polarization position angles and the polarization average profile, respectively. The profile is an average of approximately 2,000 rotations. The bottom three panels are the three single pulses blindly detected in our search as discussed in Section~\ref{ssec:sglsearch}. The solid line in the top panel stands for the RVM fit to the position angles. The solid black, dashed red and dotted blue lines in the rest of the panels represent total intensity, linear and circular component, respectively. The intensities are normalized with respect to the peak of the average profile in the second panel from the top. \label{fig:magprofgmva2017}}
\end{figure}

The 2017 observations offer a great opportunity to study the emission property of the GC magnetar, in particular its polarization. Figure~\ref{fig:magprofgmva2017} shows the polarization  profile of the magnetar, together with its linear polarization position angle ($\Psi$) swing and the three individual pulses detected from the single-pulse search described in Section~\ref{ssec:sglsearch}. It can be seen that the magnetar pulse profile is nearly 100\% linearly polarized and the circular polarization component is also clearly seen. The high degree of linear polarization was already indicated by previous observations with the IRAM 30-m telescope at 3-mm wavelength \citep{tde+17}. To study the emission geometry of the GC magnetar, we fitted the rotating-vector model (RVM) to the obtained position angle ($\Psi$) swing as described in \cite{rc69a}\footnote{Here we use the usual IAU convention where measured position angle increases counterclockwise on the sky, as discussed in \cite{ew01}}:
\begin{eqnarray} \label{eq:rvm}
&&\tan(\Psi_0-\Psi) =
\nonumber \\
&&\frac{\sin\alpha\sin(\phi-\phi_0)}{\sin(\alpha+\beta)\cos\alpha-\cos(\alpha+\beta)\sin\alpha\cos(\phi-\phi_0)},
\end{eqnarray}
where $\phi$ is the pulse phase, $\alpha$ is the magnetic inclination angle and $\beta$ is the impact parameter (angle between the magnetic axis and line of sight). The best-estimated values are: $\alpha=110_{-18}^{+16}$\,deg, $\beta=-20.0_{-1.7}^{+3.0}$\,deg (1-$\sigma$ confidence interval). We also derived the mean flux density of the GC magnetar to be $0.39\pm0.01$\,mJy based on the radiometer equation (detailed in Section~\ref{ssec:lim}), up to an order of magnitude weaker than previously reported at 3-mm wavelength \citep{tek+15,tde+17}. This is however in general consistent with measurements from some of the monitoring campaign with IRAM 30-m telescope (Torne et al. 2021 submitted). 

From Figure~\ref{fig:magprofgmva2017}, it can be seen that the three single pulses have peak intensities approximately 10-30 times of the average over all pulses. They individually show a high fraction in linear polarization as is also observed in the integrated profile. The single pulses occur at phases corresponding to the trailing part of the integrated profile. Sub-pulse structure is seen in all three cases, as also reported from previous single-pulse studies of the GC magnetar at lower frequencies \citep{pmp+18,wcc+19}.

\section{Survey sensitivity} \label{sec:sensi}
\subsection{Limitation by noise} \label{ssec:lim}
For a given pulsar survey, its sensitivity can be theoretically derived with the radiometer equation. With a requested S/N of detection, the minimum detectable mean flux density of the pulsar can be approximated by \citep{lk05}\footnote{Note that this formula is a valid approximation only when $P\gg W$. See \citet[e.g.,][]{cc97,lbh+15} for extended discussions on this topic and more generic expressions.}:
\begin{equation}
     S_{\rm min}=\frac{(\rm S/N)T_{\rm sys}}{G\sqrt{n_{\rm p}t_{\rm int}\Delta\nu}}\sqrt{\frac{W}{P-W}},
\end{equation}
where $T_{\rm sys}$ is the system temperature, $G$ is the telescope gain, $n_{\rm p}$ is the number of polarizations, $t_{\rm int}$ is the integration time, $\Delta\nu$ is the bandwidth, $P$ is the pulsar rotational period and $W$ is the effective pulse width. For our observation at Band-3, we  use $G=1.15$\,K/Jy and $T_{\rm sys}=51$\,K as a result from the QA2 analysis which also takes into account the phasing efficiency of the array \citep{gmm+19}. However, sensitivity estimates using the formula above would have a few caveats when applying to our data. First, it is known to overestimate the sensitivity for slow pulsars (e.g., $P>1$ s) due to the presence of red noise in the data \citep{lbh+15}. Secondly, the data are effectively not evenly sampled as a consequence of the gaps between individual scans. In addition, there could be impact on the sensitivity by other (periodic) signals or non-Gaussianity in the data. Thus, similar to \cite{lbh+15}, we estimated the sensitivity of our search using a few different schemes described as follows. Firstly, we directly calculated the values with the radiometer equation. Secondly, we simulated a fake pulsar signal along with white-noise data of the same length, spacing and RMS as the real observations, and carried out a real search with our pipeline. Furthermore, we directly injected a fake pulsar signal into the real data and performed a search with our pipeline. 

\begin{figure}
\centering
\hspace*{-0.4cm}
\includegraphics[scale=0.6]{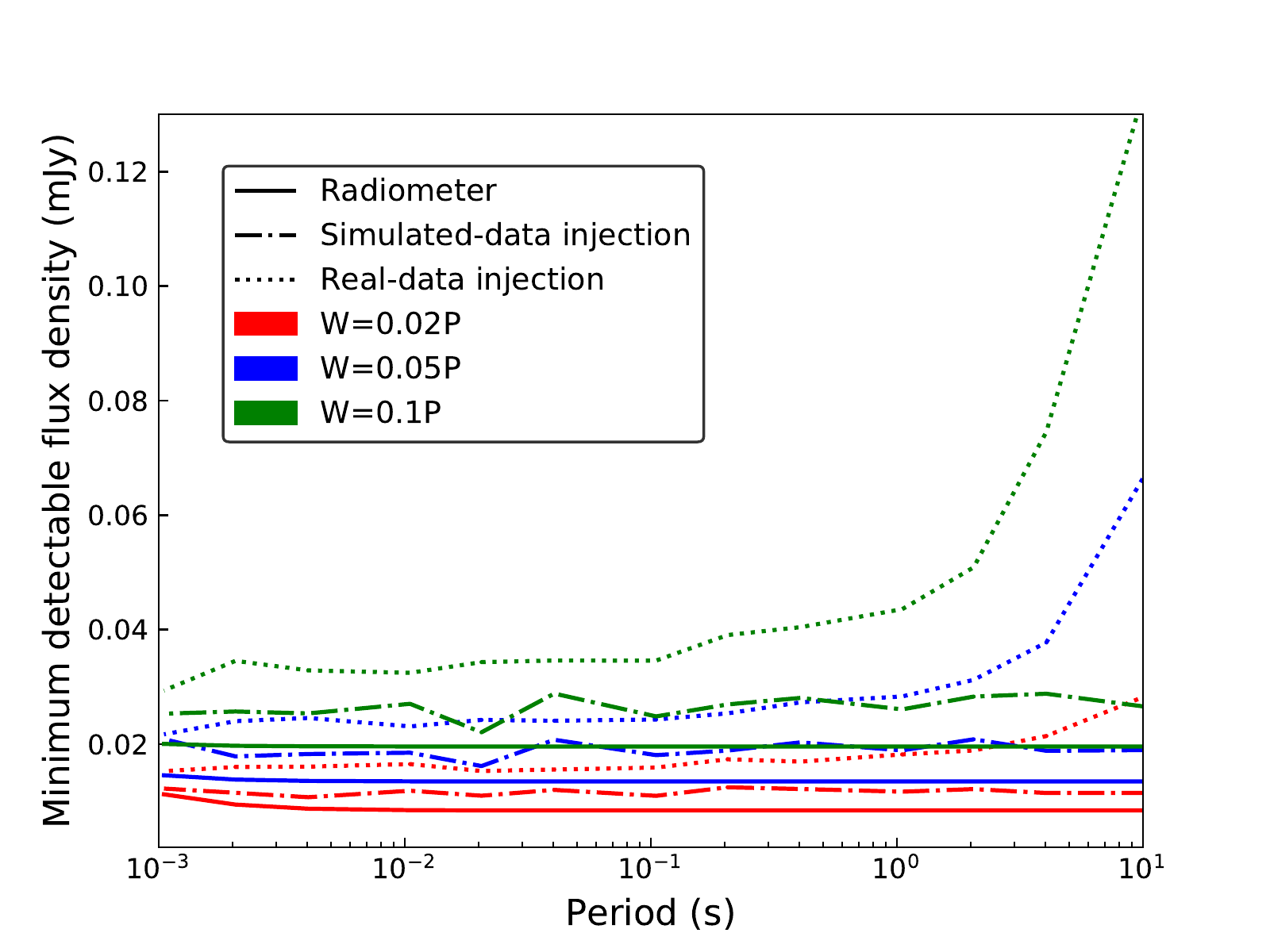}
\caption{Minimum detectable flux density (with 7-$\sigma$ detection threshold) of the GMVA survey as a function of pulsar period, estimated with the radiometer equation, injection into simulated data, and injection into the real data, respectively. Limits are shown for three typical values of pulse width ($W$). The pulse width of PSR~J1745$-$2900 is approximately $0.1\,P$ from the GMVA 2017 observation. 
\label{fig:limit}}
\end{figure}

Figure~\ref{fig:limit} summarizes the sensitivity estimates described above. It can be seen that the estimates from  radiometer equation and from injection into simulated data lead to consistent minimum detectable flux densities, while values from the latter are approximately 20\% higher possibly due to the existence of gaps in between individual scans. Injection into real data resulted in slightly higher limits for fast spinning pulsars, which is around 0.03\,mJy, and became up to a factor of 5 higher for slow pulsars ($P\gtrsim1$\,s) with wide pulse width. This is expected due to the presence of red noise and other terrestrial signals in the real data. 

\subsection{Sensitivity to binary motion} \label{ssec:bnry}

\begin{figure}
\hspace*{-0.5cm}
\includegraphics[scale=0.6]{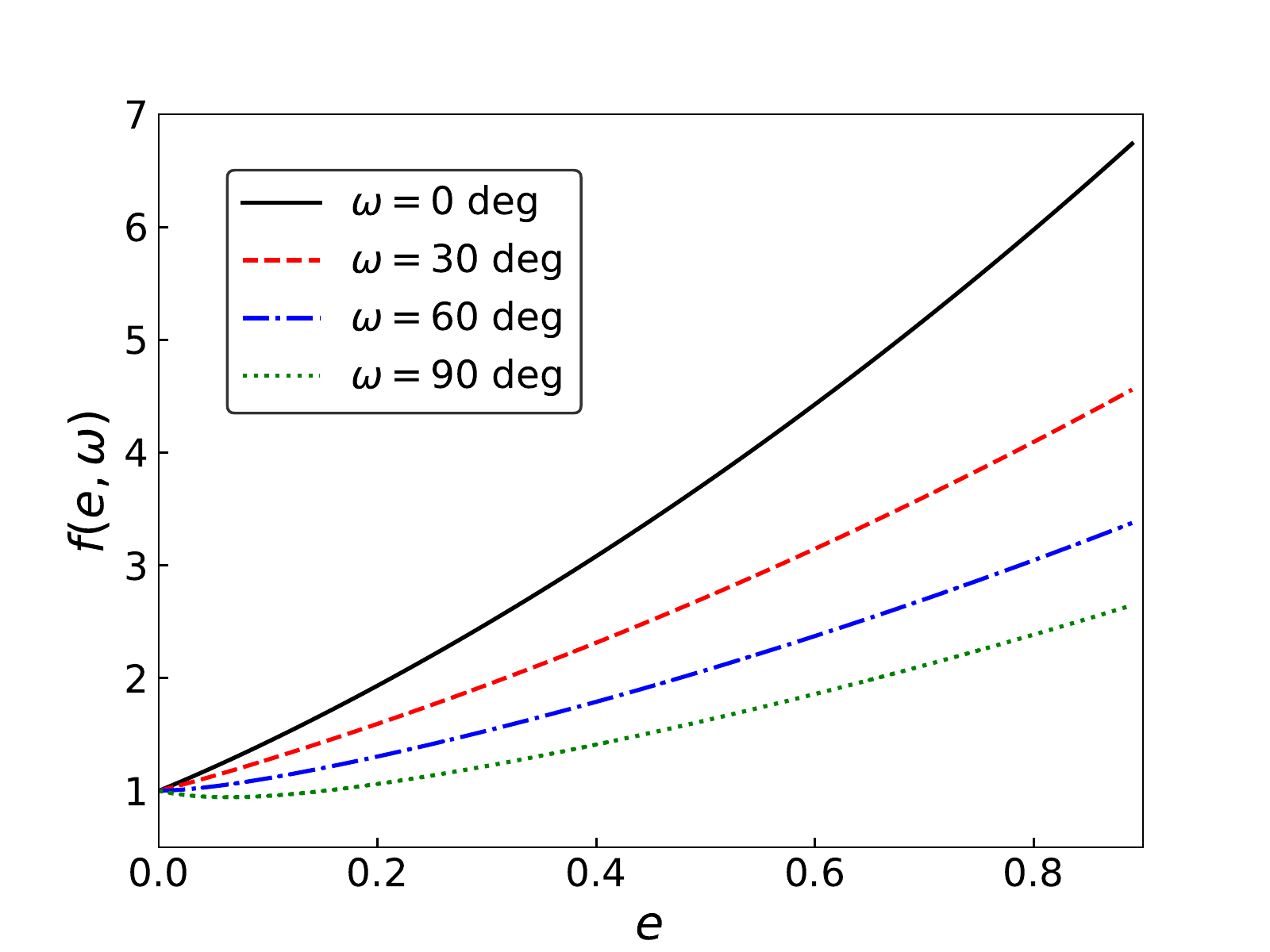}
\caption{Value of $f(e,\omega)$ in Eq.~\ref{eq:zwmax_ratio} as a function of $e$ for a few options of $\omega$. In general, $f(e,\omega)$ is increased from unity when the orbit becomes more eccentric, while having a local minimum of approximately 0.94 at $e\sim0.1$ when $\omega$ is close to 90\,deg. For a given $e$, $f(e,\omega)$ is at maximum when $\omega=0$ or $180$\,deg. The value range of $f(e,\omega)$ is identical when $\omega\in [0, 90)$, $[90, 180)$, $[180, 270)$ and $[270, 360)$\,deg, respectively. \label{fig:fe}}
\end{figure}

\begin{figure*}
\gridline{\hspace*{-0.7cm}\fig{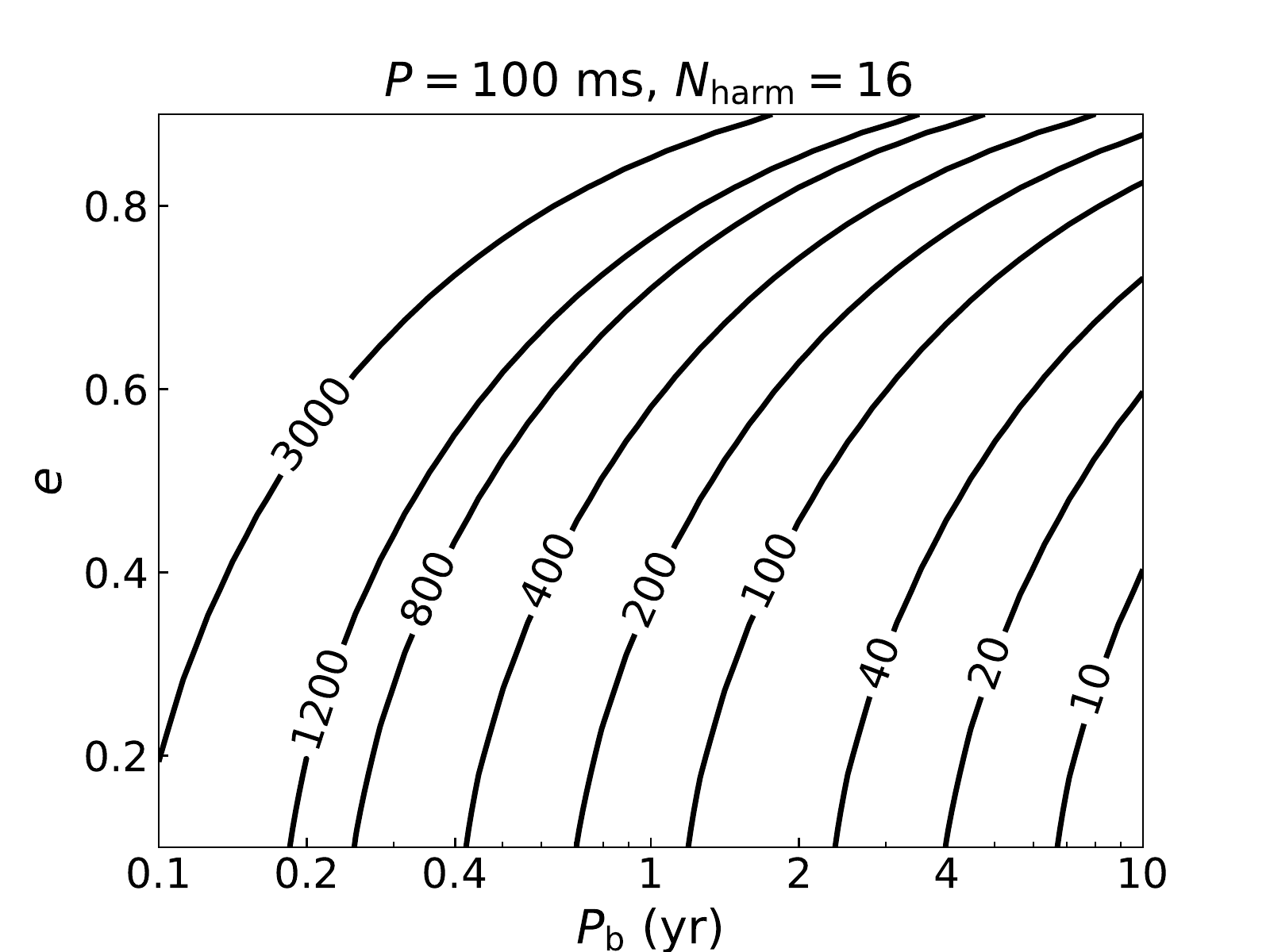}{0.285\textwidth}{}
          \hspace*{-0.5cm}\fig{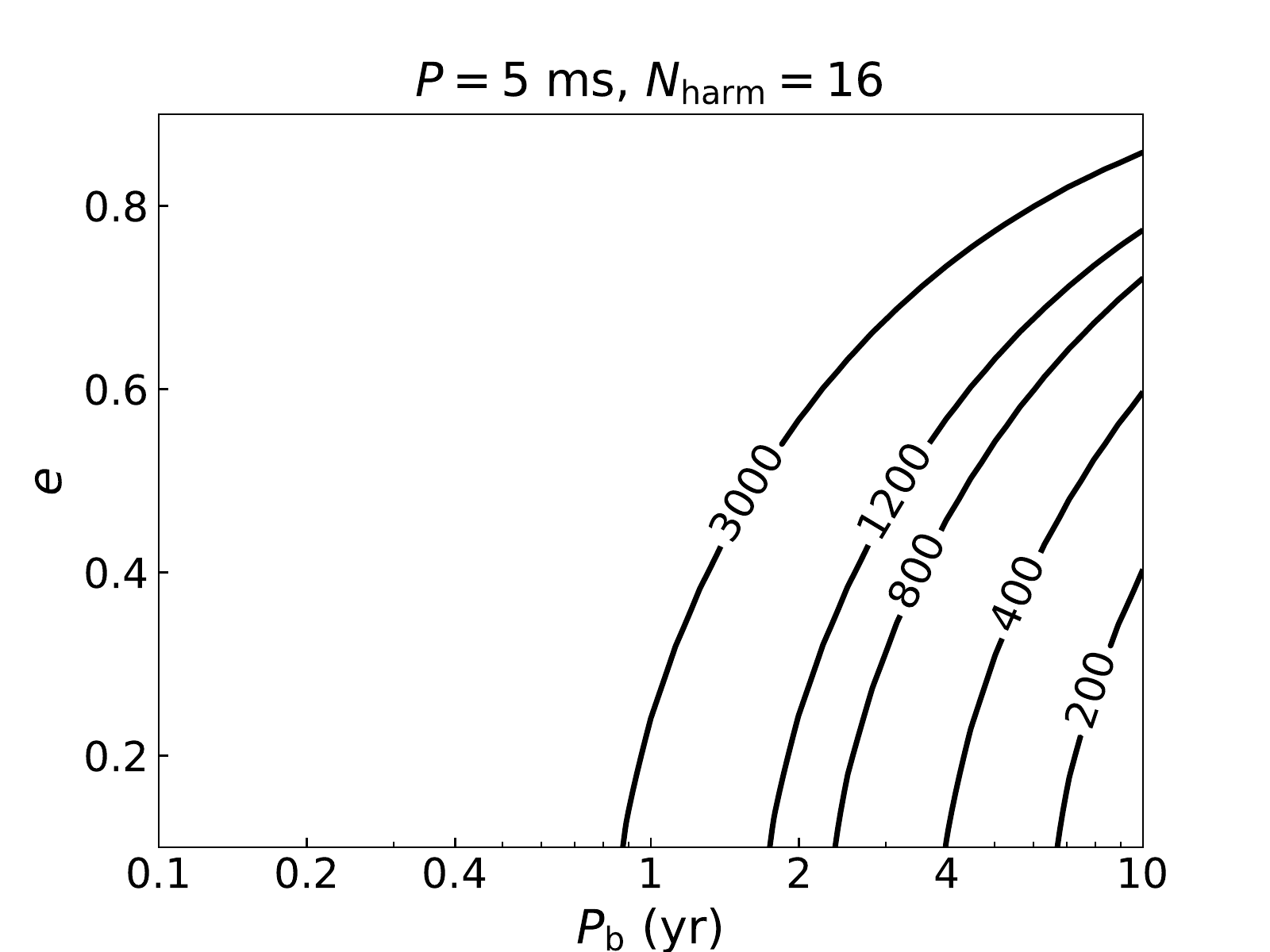}{0.285\textwidth}{}
          \hspace*{-0.5cm}\fig{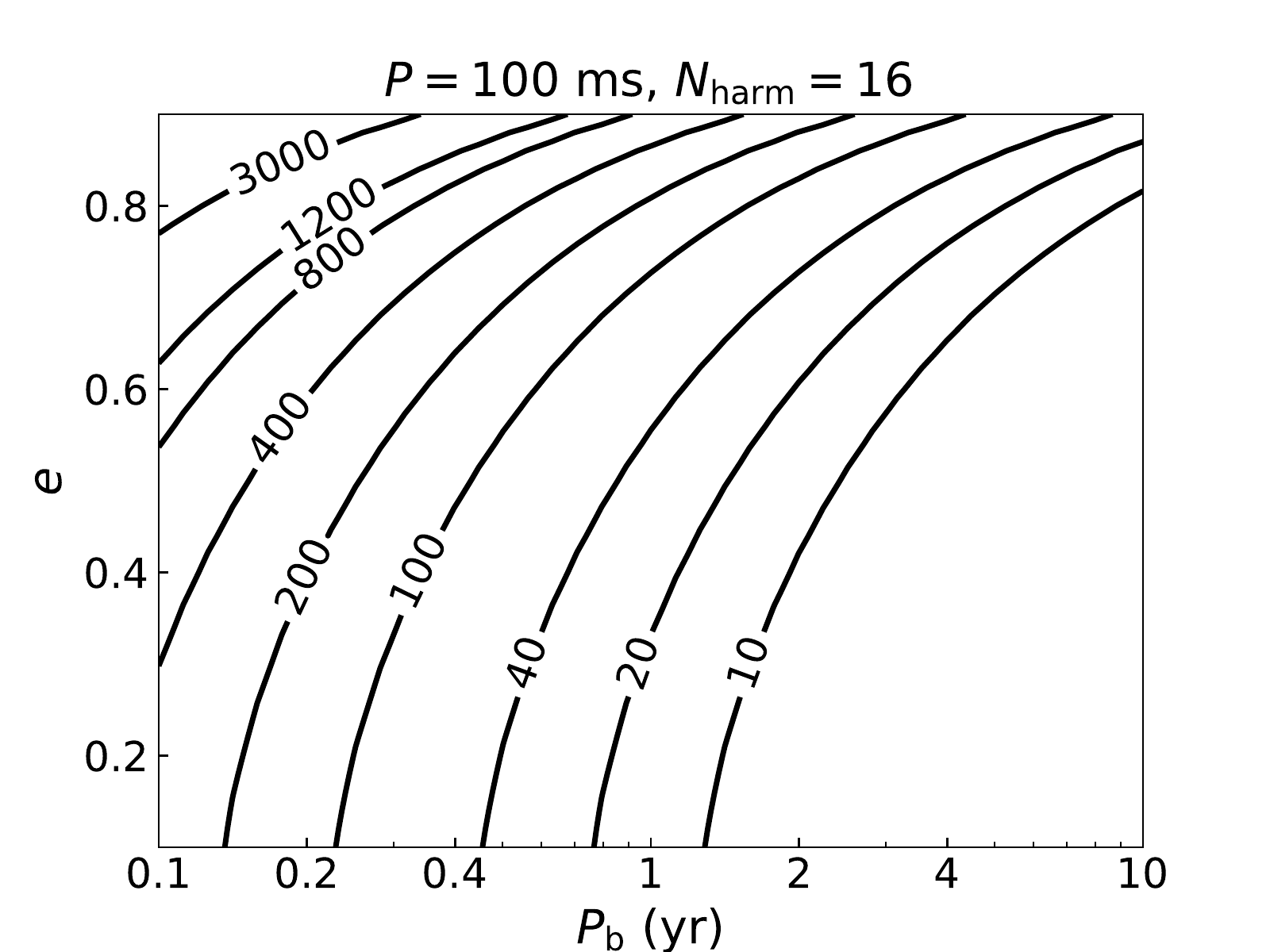}{0.285\textwidth}{}
          \hspace*{-0.5cm}\fig{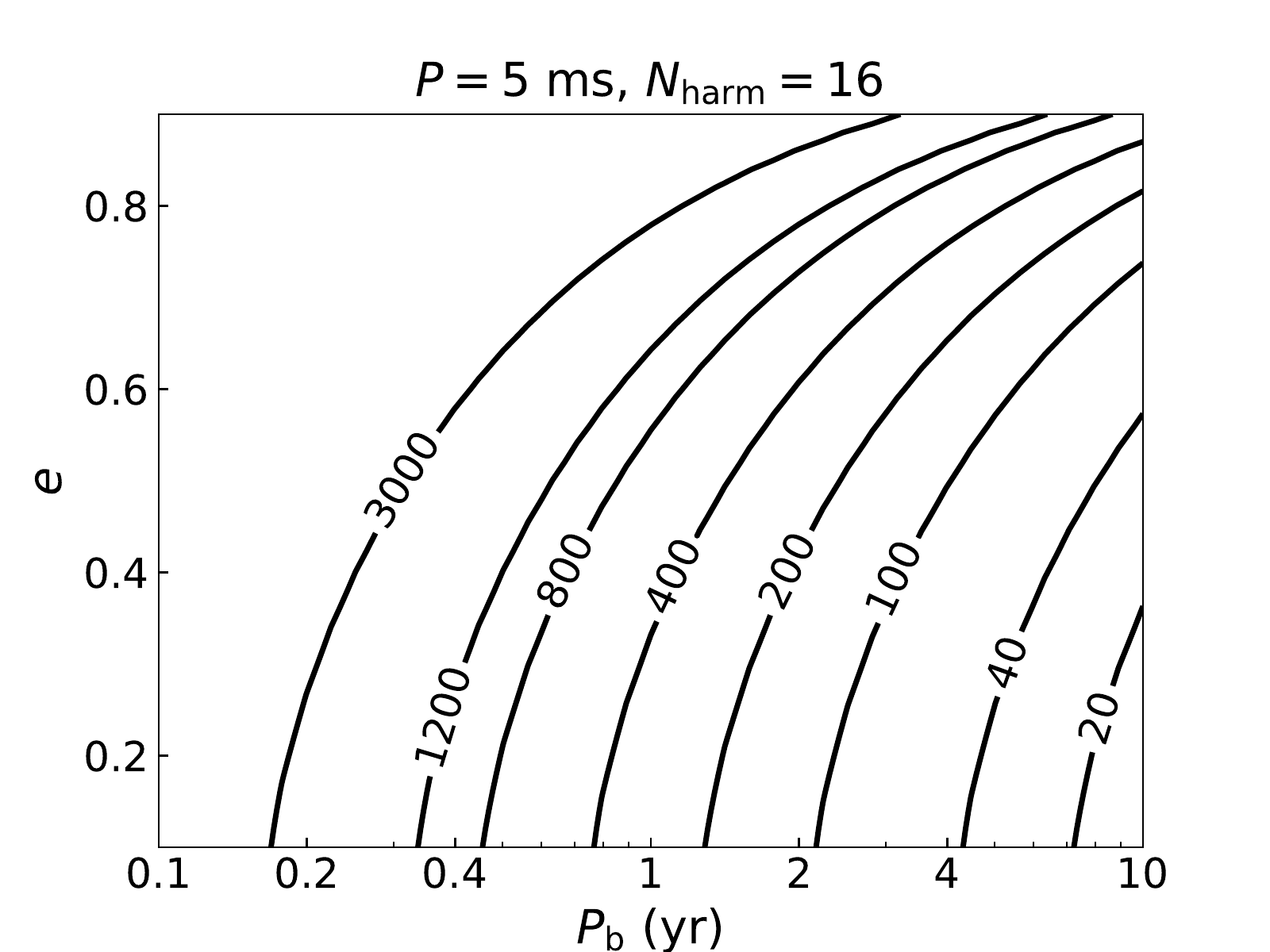}{0.285\textwidth}{}}
\vspace*{-0.9cm}
\gridline{\hspace*{-0.7cm}\fig{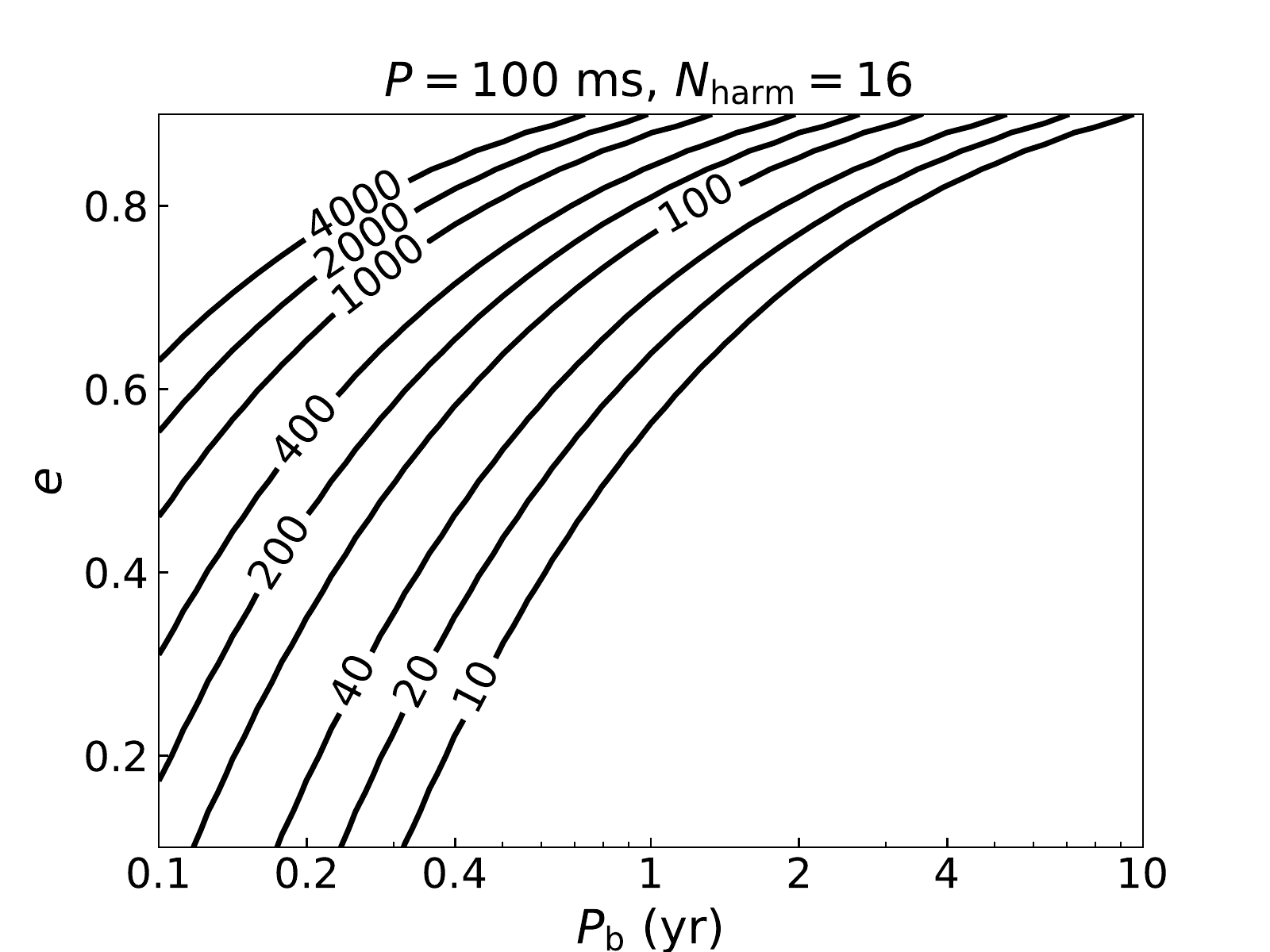}{0.285\textwidth}{Ordinary pulsar, full length}
          \hspace*{-0.5cm}\fig{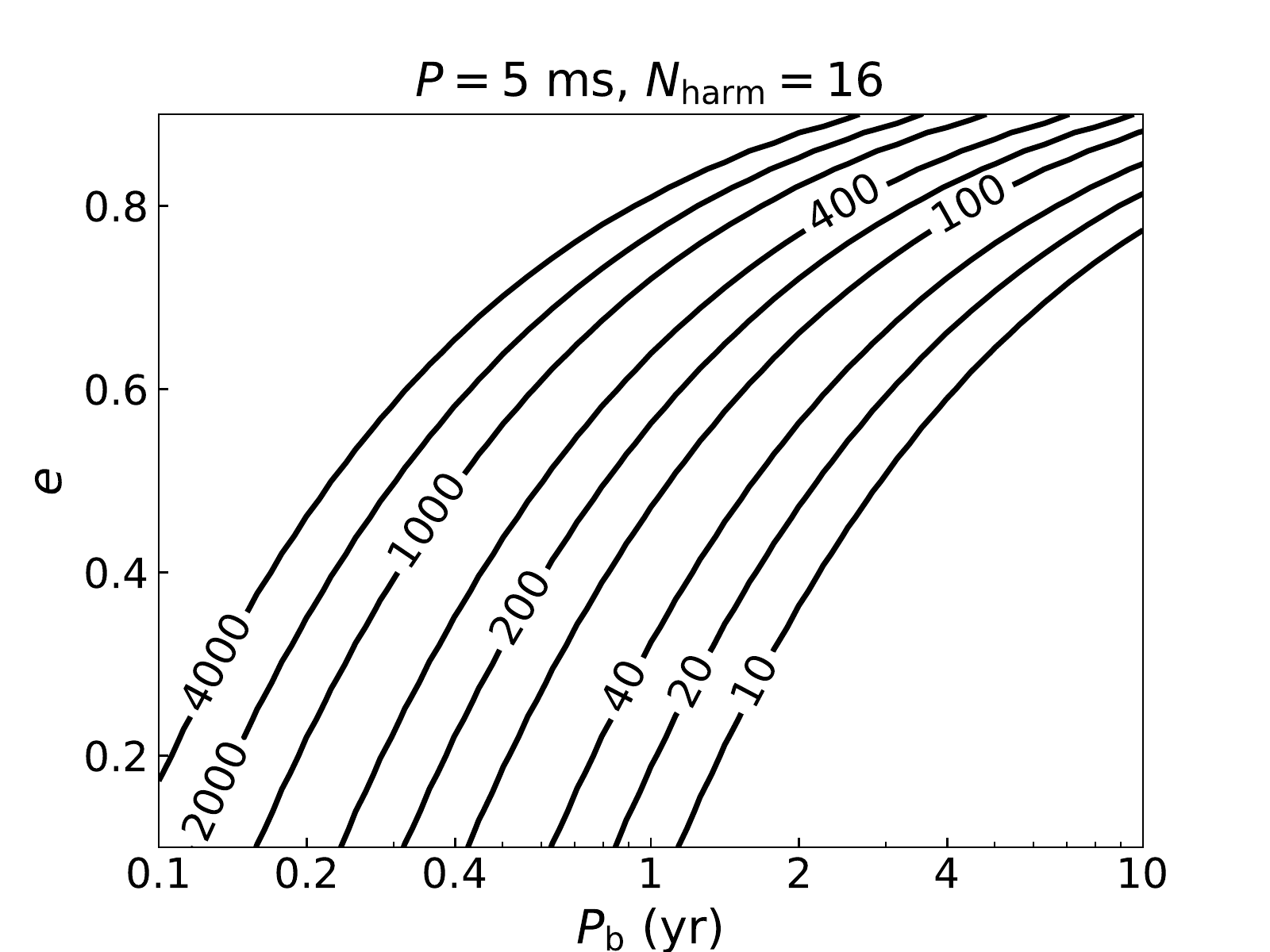}{0.285\textwidth}{MSP, full length}
          \hspace*{-0.5cm}\fig{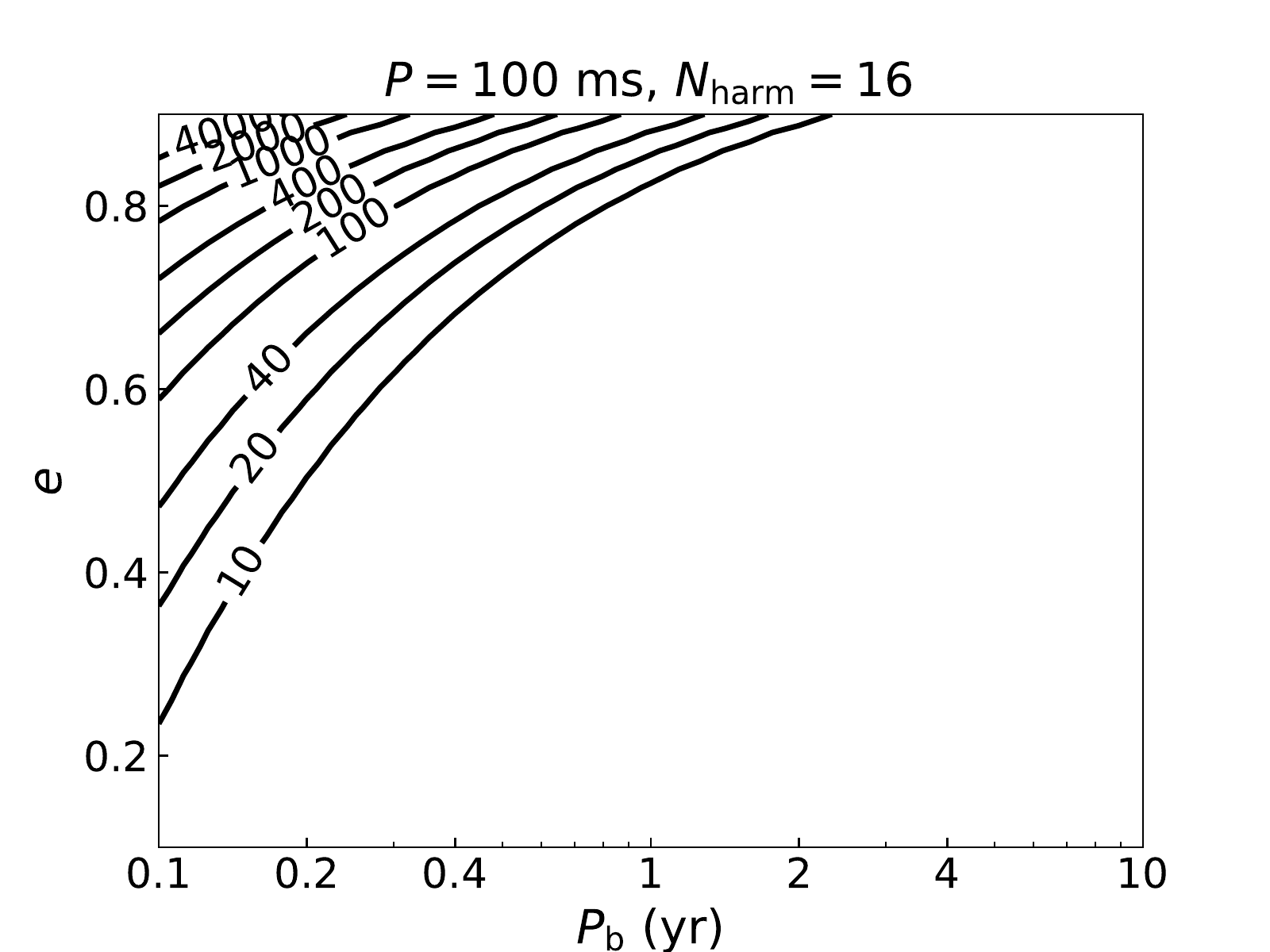}{0.285\textwidth}{Ordinary pulsar, 1/3 length}
          \hspace*{-0.5cm}\fig{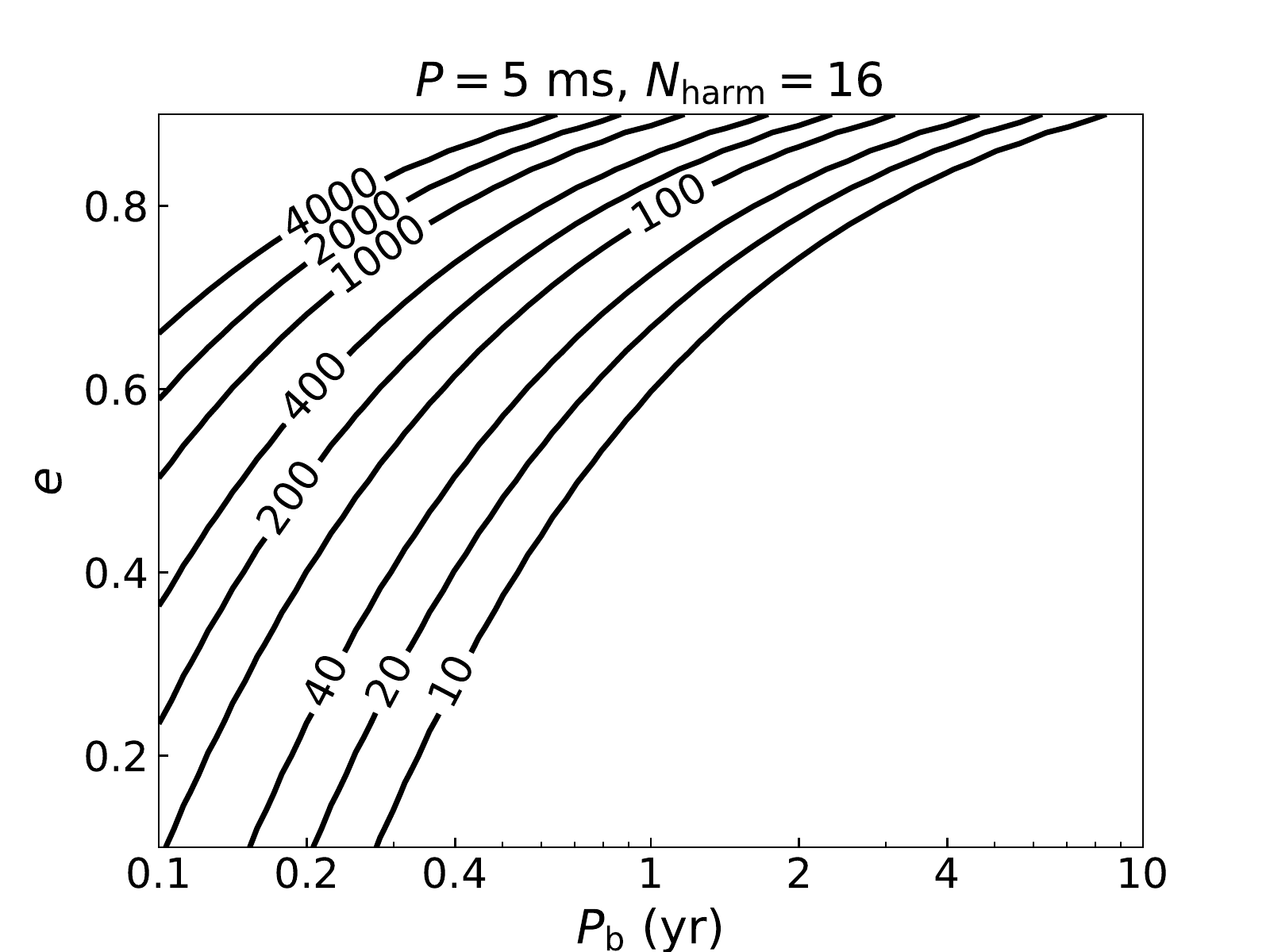}{0.285\textwidth}{MSP, 1/3 length}}
\caption{\texttt{zmax} (upper row) and \texttt{wmax} (lower row) values required in \textsc{presto} for our searches to retain optimal sensitivity to pulsars in orbit with Sgr~A*, for a range of orbital periods and eccentricities. Here we use the length of the 2017 campaign (5.1\,hr in total) and the 2018 campaign has a similar length. As mentioned in Section~\ref{ssec:psearch}, the \texttt{zmax}, \texttt{wmax} values used in our searches are 1200, 40 for the full data length, and 1200, 1500 for the 1/3 data length, respectively.
\label{fig:accel_sgrA}}
\end{figure*}

The observed period of a pulsar can change significantly during the observation as a result of the Doppler effect if it is in orbital motion and thus accelerated. The variation depends on the orbital phase at the epoch of observation, and is approximately linear in time if the pulsar endures roughly a constant line-of-sight acceleration ($a_{\rm l}$) during the observation \citep{rem02}. However, if the observing length ($T_{\rm obs}$) is a significant fraction of orbital period (e.g., $T_{\rm obs}\gtrsim10\%~P_{\rm b}$), the derivative of the acceleration, commonly referred to as jerk ($j_{\rm l}$), is then required to better describe the variation of pulsar period \citep{rem02,blw13,ekl+13,lew+14,ar18}. While acceleration and jerk searches have been widely studied in searches for binary pulsars with solar-mass companions, important differences when searching for pulsars in orbit with Sgr A* have only been pointed out very recently by Eatough et al. (2021, submitted). Here, it was shown that in deep searches for millisecond pulsars (MSPs) in long-period circular orbits around Sgr~A* $P_{\rm b}\sim2$\,yr, even the constant acceleration is enough to make $z$ exceed the current maximum value configurable in {\sc presto accelsearch} $-$ thereby placing lower limits on the minimum Sgr~A* orbital period detectable, for a pulsar of a given spin period.

Following e.g., \cite{blw13}, the line-of-sight acceleration and jerk of an orbital pulsar can be written as:
\begin{equation}
    a_{\rm l}=-\left(\frac{2\pi}{P_{\rm b}}\right)^2\frac{a_{\rm p}\sin i}{(1-e^2)^2}\sin(A_{\rm T}+\omega)(1+e\cos A_{\rm T})^2,
\end{equation}
and
\begin{eqnarray}
j_{\rm l}&=&-\left(\frac{2\pi}{P_{\rm b}}\right)^3\frac{a_{\rm p}\sin i}{(1-e^2)^{7/2}}(1+e\cos A_{\rm T})^3\cdot[\cos(A_{\rm T}+\omega) \nonumber\\
    &&+e\cos\omega-3e\sin(A_{\rm T}+\omega)\sin A_{\rm T}],
\end{eqnarray}
where $a_{\rm p}$, $e$, $i$, and $A_{\rm T}$ are the semi-major axis, eccentricity, inclination and true anomaly of the pulsar orbit. In the Fourier domain, $a_{\rm l}$ and $j_{\rm l}$ correspond to shifts in the Fourier frequency bin ($z$) and its derivative ($w$) within an entire observation of length $T_{\rm obs}$ as \citep{ar18}
\begin{equation}
    z=\frac{a_{\rm l}hfT_{\rm obs}^2}{c},
\end{equation}
and
\begin{equation}
    w=\frac{j_{\rm l}hfT_{\rm obs}^3}{c},
\end{equation}
where $c$, $f$, $h$ are the speed of light, fundamental frequency of the pulsar rotation, and the index of the Fourier harmonic being considered. In practice, the Fourier-domain search is blindly performed, covering a range of possible $z$ and $w$ values, respectively. Each range is symmetric around zero and given a half width that covers the largest possible absolute $z$ or $w$ from an entire orbit. While in reality the absolute $z$ and $w$ reach their maximum at different orbital phase, i.e., observing epochs, a blind search needs to cover both maxima at the same time so as to ensure an optimal survey sensitivity at any potential orbital phase. With the equations above, the ratio of the maximum of absolute $z$ and $w$ out of an entire orbit can be written as
 \begin{equation} \label{eq:zwmax_ratio}
    \frac{{\rm max}(|w|)}{{\rm max}(|z|)} = \frac{T_{\rm obs}}{P_{\rm b}(1-e^2)^{3/2}}f(e,\omega),
\end{equation}
where $f(e,\omega)=1$ for $e=0$, and in general increases for a larger $e$ as shown in Figure~\ref{fig:fe}. It can be seen that the relative value of ${\rm max}(|w|)$ to ${\rm max}(|z|)$ becomes larger for more compact and eccentric orbits, and longer observations. This shows the increasing importance of including a jerk search under these circumstances. 

Accordingly, for a given observation we can calculate the maximum absolute $z$ and $w$ values from a range of orbits, to see if the search is in practice sensitive to pulsars in those orbits. Here for the estimate of our search, we mainly consider two scenarios: 1) A pulsar in orbit with Sgr~A*; 2) A pulsar in a close binary with a degenerate companion (a neutron star or white dwarf). We also consider the impact on the detectability by orbital eccentricity where jerk effects (covered by the $w$ term) become more significant, in comparison with the analysis presented in Eatough et al. (2021, submitted). The results for our survey concerning the first scenario are presented in Figure~\ref{fig:accel_sgrA}. It can be seen that for an ordinary pulsar ($P=100$\,ms), a maximum $|z|$ value of 1200 is sensitive to orbital periods down to $\sim0.5$\,yr for a moderate eccentricity ($e\sim0.5$), and further down to $0.2$\,yr if the eccentricity is low ($e\sim0.1$). For a MSP, these are restricted to $\sim5$\,yr and $2$\,yr, respectively. The searches that use 1/3 of the entire observation and the same maximum $|z|$, significantly improves the coverage of the orbital parameters space. For an ordinary pulsar, it is expected to be sensitive to the vast majority of the possible orbits except for those of a very short orbital period and high eccentricity. For an MSP, the segmented search begins to cover a significant fraction of possible orbits with $P_{\rm b}<1$\,yr. Exploration in the space of jerk in this scenario is much less required except only for short and highly eccentric orbits, as indicated from the scaling law by Eq.~(\ref{eq:zwmax_ratio}). For a choice of ${\rm max}(|z|)=1200$, a corresponding ${\rm max}(|w|)=40$ used in our search is just enough to cover the same range of orbits while using the full length of the observation. For the segmented searches that use 1/3 of the full length, the required ${\rm max}(|w|)$ is approximately 400 which is well below 1500 which we used. Our results are in broad consistency with what have been reported by Eatough et al. (2021 submitted) (shown in Figure 7), when the orbital eccentricity is low ($e=0.1$).

\begin{figure}
\gridline{\hspace*{-0.7cm}\fig{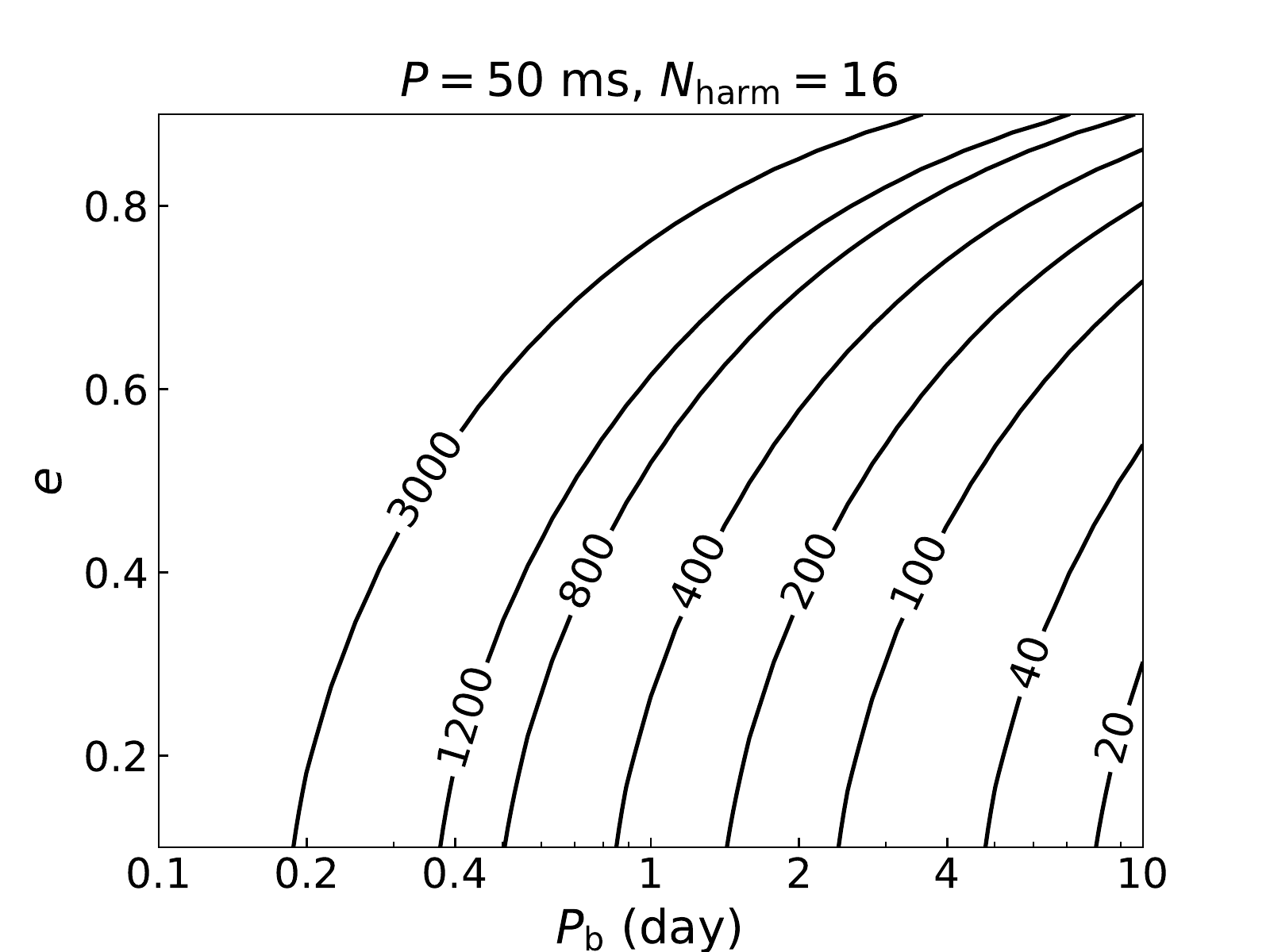}{0.285\textwidth}{}
          \hspace*{-0.5cm}\fig{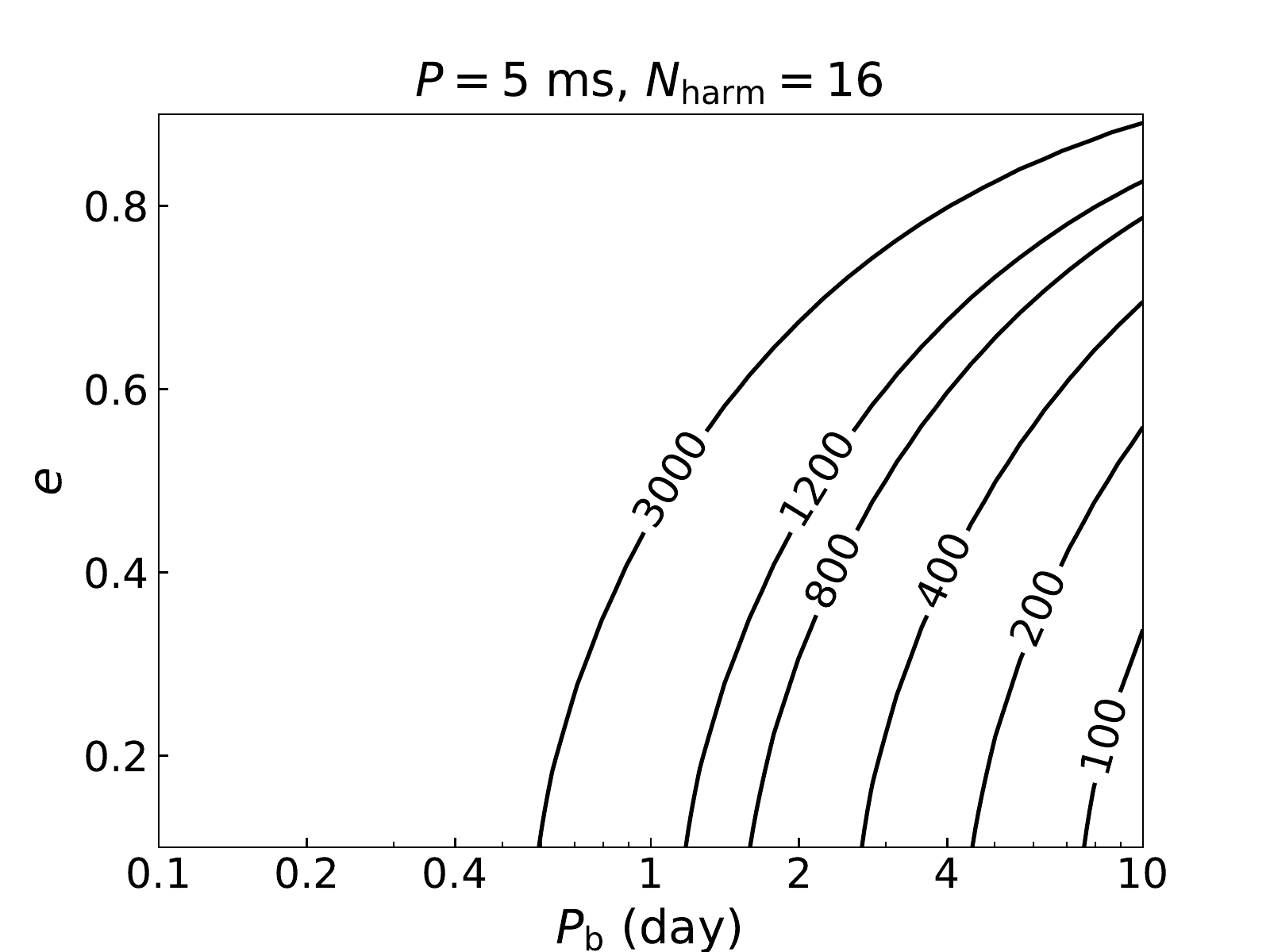}{0.285\textwidth}{}}
\vspace*{-0.9cm}
\gridline{\hspace*{-0.7cm}\fig{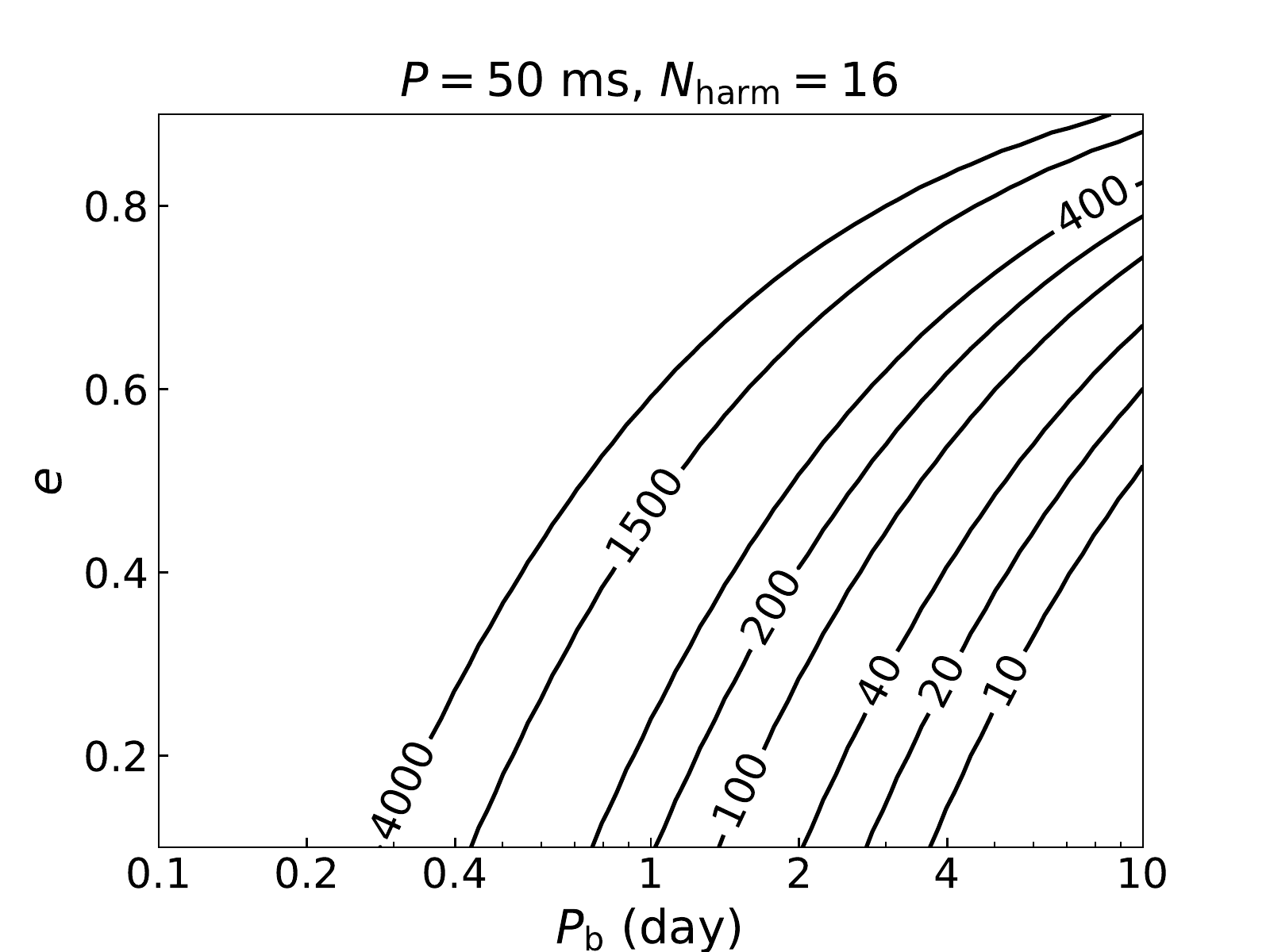}{0.285\textwidth}{}
          \hspace*{-0.5cm}\fig{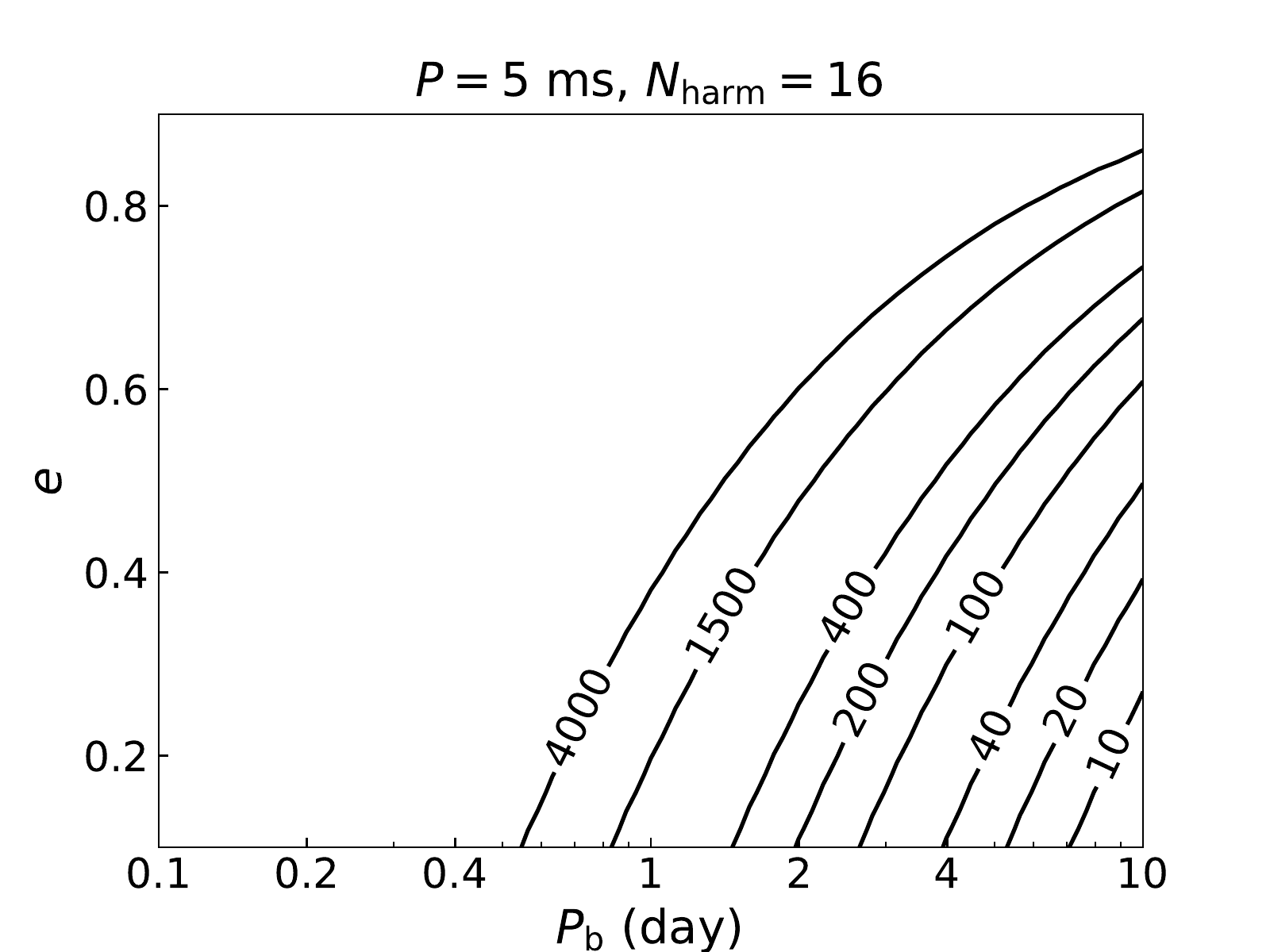}{0.285\textwidth}{}}
\caption{Maximum of $z$ (upper row) and $w$ (lower row) values required in our segmented search (1/3 length) for orbits of different periods and eccentricities in a binary pulsar system. The companion is assumed to be a 1.4-M$_{\odot}$ neutron star for the case of a mildly recycled pulsar ($P=50$\,ms), and a 0.5-M$_{\odot}$ white dwarf for the case of a MSP ($P=5$\,ms). The full observing length used here is the same as in Figure~\ref{fig:accel_sgrA}. For reference, the \texttt{zmax}, \texttt{wmax} values used in the 1/3-length searches are 1200 and 1500, respectively, as mentioned in Section~\ref{ssec:psearch}. \label{fig:accel_bnry}}
\end{figure}

For many of the known binary pulsars, our 5.2-hr long dataset is comparable to or larger than the orbital period 
\citep{mhth05}. Thus, the segmented search is necessary to provide sensitivity for a significant population of binary pulsars, in particular those with $P_{\rm b}\lesssim1$\,d. Figure~\ref{fig:accel_bnry} shows the maximum $z$ and $w$ values required in the segmented search for a range of orbital period and eccentricity. Here two common types of binary pulsars are considered, a mildly recycled pulsar with a neutron star companion (1.4\,M$_{\odot}$) and an MSP with a white dwarf companion (0.5\,M$_{\odot}$) \citep{lk05}. It can be seen that with maximum $z$ of 1200 and maximum $w$ of 1500 used in our search, the coverage of orbit is approximately down to 0.7\,d in the former and 2\,d in the latter case, for a moderate orbital eccentricity ($e\sim0.5$). These constraints drop to 0.5\,d and 1\,d, respectively, if the eccentricity is assumed to be low ($e\sim0.1$).

The aforementioned sensitivity limit of our search on potential pulsar orbits should be considered as constraints on the ``worst-case scenario'', since it requires the survey to be conducted with an ``optimal'' sensitivity. In practice, a pulsar in a tight orbit, where a jerk search or a higher order derivative of acceleration 
is needed to approximate the orbital motion, could still be detected if its signal is sufficiently strong after the systematic reduction. In the Fourier domain, the amplitudes of   
high-order harmonics of the pulsar rotational frequency are reduced, which causes harmonic summing to also produce a smaller detection statistic.
It is also possible that the pulsar happens to be observed at a favourable orbital phase, where the instantaneous values of $z$ and $w$ are below the limits while their maxima over an entire orbit are not. Lastly, single pulses are unaffected by imperfect removal of the orbit from arrival times. 

\section{Discussion} \label{sec:dis}

\begin{figure}
\centering
\hspace*{-0.4cm}
\includegraphics[scale=0.6]{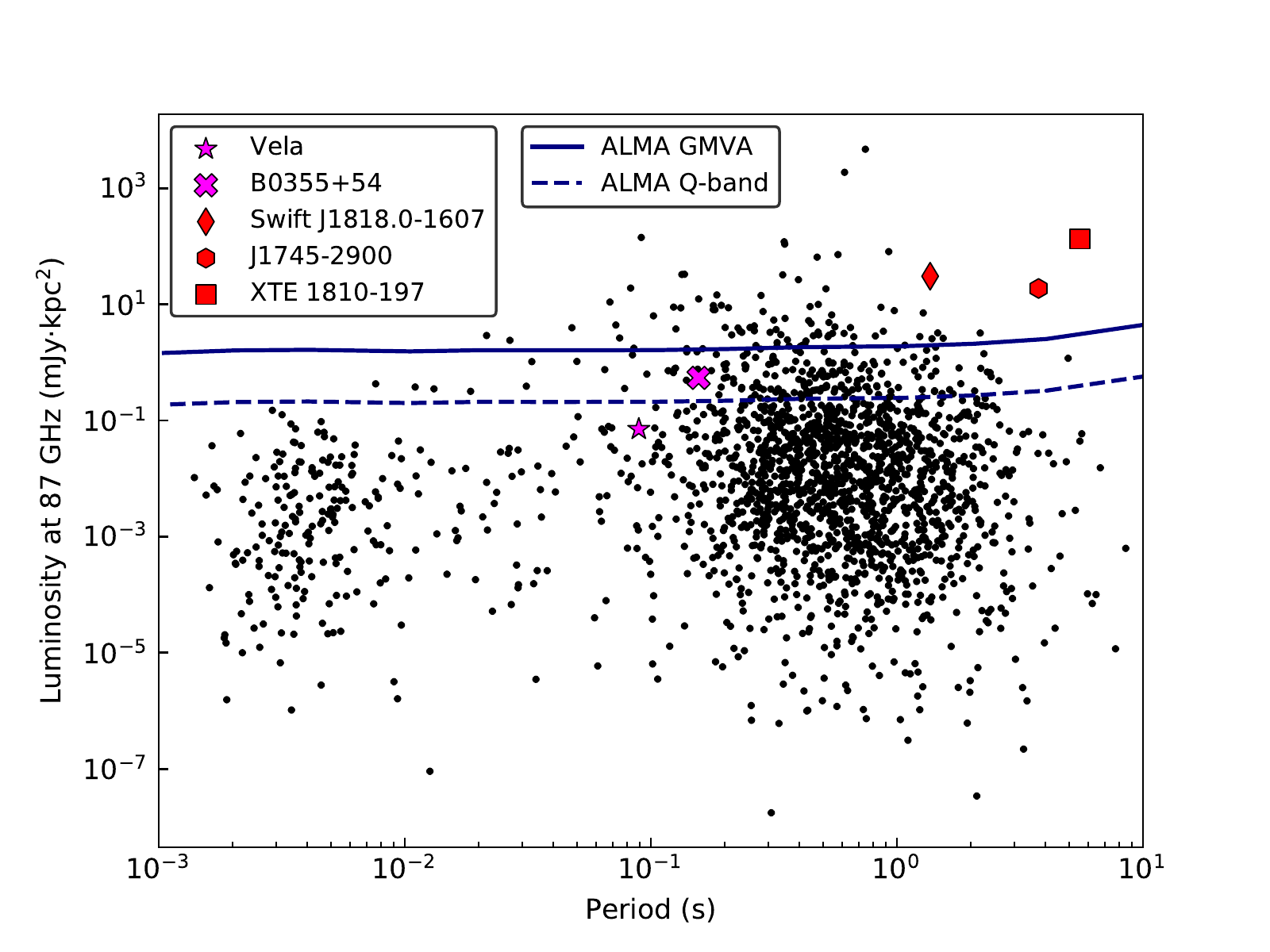}
\caption{Sensitivity of ALMA GC surveys with respect to the 87-GHz luminosity of a large pulsar population. The black dots represent luminosities calculated from one typical iteration out of a thousand in each of which a random spectral index was drawn from a normal distribution for pulsars without reported spectral index measurement. The ALMA GMVA sensitivity curve is obtained from the injection into real data with $W=0.05$\,$P$ shown in Figure~\ref{fig:limit}. Five pulsars that have been detected at mm-wavelength are highlighted, where the purple and red color represent ordinary pulsar and magnetar, respectively. 
\label{fig:PvsL}}
\end{figure}

It is anticipated that the GC hosts a substantial population of pulsars in its inner parsec, with the most optimistic estimate of up to $\sim100$ pulsars\footnote{This is merely a population estimate and does not take into account the beaming fraction of pulsars, roughly 20\% estimated from previous observations \citep[e.g.,][]{lk05}.} with $P_{\rm b}\lesssim10$\,yr that can be covered by our search \citep[e.g.,][]{pl04,wcc+12,cl14}. However, as discussed in \cite{elc+15} and \cite{le17}, most surveys by now are in practice very limited in exploring the GC pulsar population. To examine how deeply our survey can probe the GC pulsars, we have calculated the luminosity threshold at 87\,GHz based on the estimated survey sensitivity in Section~\ref{sec:sensi}. As a comparison, we also estimated the anticipated sensitivity of a Q-band survey with ALMA, once its Band-1 is fully operational. For this we applied a similar observing setup (6-hr tracking, effective on-source time approximately 50\%) to the GMVA campaign, 8-GHz bandwidth (with all four spectral windows), the same telescope gain as in Band-3, a system temperature of 50\,K and a central frequency of 42\,GHz \citep{hmk+16}. For a given pulsar period, the system sensitivity was then derived based on the estimate in Figure~\ref{fig:limit} with the radiometer equation, and translated into a luminosity threshold at 87\,GHz assuming a spectral index of $-1.6$.

These limits were compared with the anticipated luminosities of a large pulsar population. To select the samples of pulsars, we used the ATNF pulsar catalogue \citep{mhth05}, and chose all pulsars with flux density measurements around 1.4\,GHz or above, because the pulsar spectrum at lower frequencies tends to 
deviate from a simple power-law \citep[e.g.,][]{klm+11,jvk+18}. This kept 70\% (1,932 out of 2,796) of the overall pulsar population for the analysis thereafter. Then for pulsars with a reported spectral index, 
we extrapolated the flux density to 87~GHz assuming the power-law spectrum extends up to that frequency. If no measurement of spectral index is available, we applied a random value drawn from a normal distribution (with mean of $-1.6$ and standard deviation of 0.54) reported in \cite{jvk+18} which was obtained based on statistics of $\sim300$ pulsars. Exceptions are PSR~J0835$-$4510 (the Vela pulsar), PSR~B0355+54, Swift\,J1818.0$-$1607, PSR~J1745$-$2900 and XTE\,J1810$-$197 for which we simply used the reported flux density measurements at 87\,GHz \citep[][and this work]{mkt+97,lyw+19,tlc+20,tml+20}. The distance measurements are mostly based on DM and the YMW16 Galactic free-electron density model \citep{ymw17}, except for the Vela pulsar, PSR~B0355+54 and XTE\,J1810$-$197 where a parallax distance was used \citep{cdmb01,ccv+04,ddl+20}, and PSR~J1745$-$2900 for which we applied the distance to the GC reported by \cite{gaa+21}. We repeated the aforementioned procedure for 1,000 realizations to accumulate a robust assessment of the pulsar population detectable by the surveys. Figure~\ref{fig:PvsL} presents the results from one typical iteration.

It has been found that our ALMA GMVA survey is able to detect only the most luminous pulsars, approximately the top $4$\% (average from all realizations) of all selected pulsars. Meanwhile, the survey would already be sensitive enough to detect all three radio-loud magnetars (Swift\,J1818.0$-$1607, PSR~J1745$-$2900 and XTE\,J1810$-$197) which have been detected at mm-wavelength\footnote{The pulse widths of Swift\,J1818.0$-$1607 and XTE\,J1810$-$197 reported in \cite{tlc+20} and \cite{tml+20}, respectively, are both close to $0.05\,P$ used in the sensitivity curve plotted in Figure~\ref{fig:PvsL}.}, if they were located in the GC. The ALMA Q-band survey has the potential to increase the fraction to approximately $14$\%, nearly a factor of 4 improvement. A similar investigation on probing the GC pulsar population with ALMA was recently reported in Torne et al., (2021 submitted), where a result in line with that presented here can be found. Note that as shown in Figure~\ref{fig:PvsL}, the surveys are very unlikely to be sensitive to any MSPs in the GC. 

The estimate above largely depends on extrapolation of pulsar spectra from decimeter and centimeter to 3-mm wavelength with a simple power-law model. This assumption has a significant number of exceptions where a more complex model, such as a broken power-law, needs to be introduced to fit the observed pulsar spectrum \citep{mkkw00,klm+11,jvk+18}. So far, only a few dozens of pulsars have had their emission properties studied above 10\,GHz \citep{wsg+72,bsw77,sw87,wjkg93,kxj+96,kjdw97,ljk+08b,kjl+11,hej16,tde+17,lyw+19}, so that an average spectral index calculated from those sources may be biased towards flatter values. On the other hand, some individual pulse components have markedly different spectra, leading to possibly complex pulsar spectra overall and rather different pulse profiles at high frequencies (see e.g.~PSR B0144$+$59, \citealt{ljk+08b}). There are also indications that the spectrum could turn up at mm-wavelengths \citep{kjdw97}. This is not completely unexpected \citep{cor79,kjdw97} as some 
theories predict the existence of other spectral components rising somewhere in between the radio and infra-red bands, e.g.~due to incoherent curvature emission \citep{mic82} or inverse Compton scattering of low-frequency radio photons \citep{bs76b,lyu13}. Therefore, additional sample studies are in great demand to fully understand pulsar emissions and the potential of pulsar surveys in the mm-wavelength regime.

\section{Conclusions} \label{sec:conc}
We have carried out the first GC pulsar and transient search with ALMA, using data collected in the GMVA 2017 and 2018 campaigns. Periodicity searches were conducted on timeseries of all four Stokes parameters, with acceleration and jerk, and in both the full length and 1/3 length of the entire observation. Single-pulse searches were performed on timeseries in total intensity, linear and circular polarization. We obtained the first polarization profile of the GC magnetar at mm-wavelength, and found that it is close to 100\% linearly polarized. While no new pulsar is found, we estimated the survey sensitivity using both the radiometer equation and a signal-injection scheme, and evaluated its capability of finding pulsars with orbital motion. Finally, we showed the survey is only sensitive to the most luminous pulsars in the overall population, and a survey using ALMA Band-1 has the potential to probe much deeper into the GC pulsar population. Future searches at mm-wavelength is necessary also to increase the detection probability of magnetars and many other pulsars of a transient nature, and pulsars at different phases of eccentric orbits. 

\acknowledgments
We thank Eduardo Ros and Marylin Cruces for carefully reading the manuscript and providing valuable suggestions. KL, GD, RPE, RK, MK, PT, RW, CG, FA, CDB, HF, AN, LR, LS acknowledge the financial support by the European Research Council for the ERC Synergy Grant BlackHoleCam under contract no. 610058. RPE is supported by a ``FAST Fellowship'' under the ``Cultivation Project for FAST Scientific Payoff and Research Achievement of the Center for Astonomical Mega-Science, Chinese Academy of Sciences (CAMS-CAS)''. JMC and SC acknowledge support from the National Science Foundation (NSF AAG-1815242). SMR is a CIFAR Fellow. RSL is supported by the Max Planck Partner Group of the MPG and the CAS and acknowledges support from the Key Program of the National Natural Science Foundation of China (grant No. 11933007) and the Research Program of Fundamental and Frontier Sciences, CAS (grant No. ZDBS-LY-SLH011). LS was supported by the National SKA Program of China (2020SKA0120300), the National Natural Science Foundation of China (11991053, 11975027), the Young Elite Scientists Sponsorship Program by the China Association for Science and Technology (2018QNRC001), and the Max Planck Partner Group Program funded by the Max Planck Society. The ALMA Phasing Project was principally supported by a Major Research Instrumentation award from the National Science Foundation (award 1126433) and an ALMA North American Development Augmentation award to Cornell University; the ALMA Pulsar Mode Project was supported by an ALMA North American Study award. 

This research has made use of data obtained with the Global Millimeter VLBI Array (GMVA), which consists of telescopes operated by the MPIfR, IRAM, Onsala, Metsahovi, Yebes, the Korean VLBI Network, the Greenland Telescope, the Green Bank Observatory and the Very Long Baseline Array (VLBA). The National Radio Astronomy Observatory is a facility of the National Science Foundation operated under cooperative agreement by Associated Universities, Inc. The VLBA is a facility of the National Science Foundation operated under cooperative agreement by Associated Universities, Inc. The data were correlated at the correlator of the MPIfR in Bonn, Germany.

\facility{ALMA, GMVA}

\software{PRESTO}

\appendix
\section{Systematics and cleaning scheme} \label{app:sys}
\begin{figure*}
\centering
\includegraphics[scale=0.4]{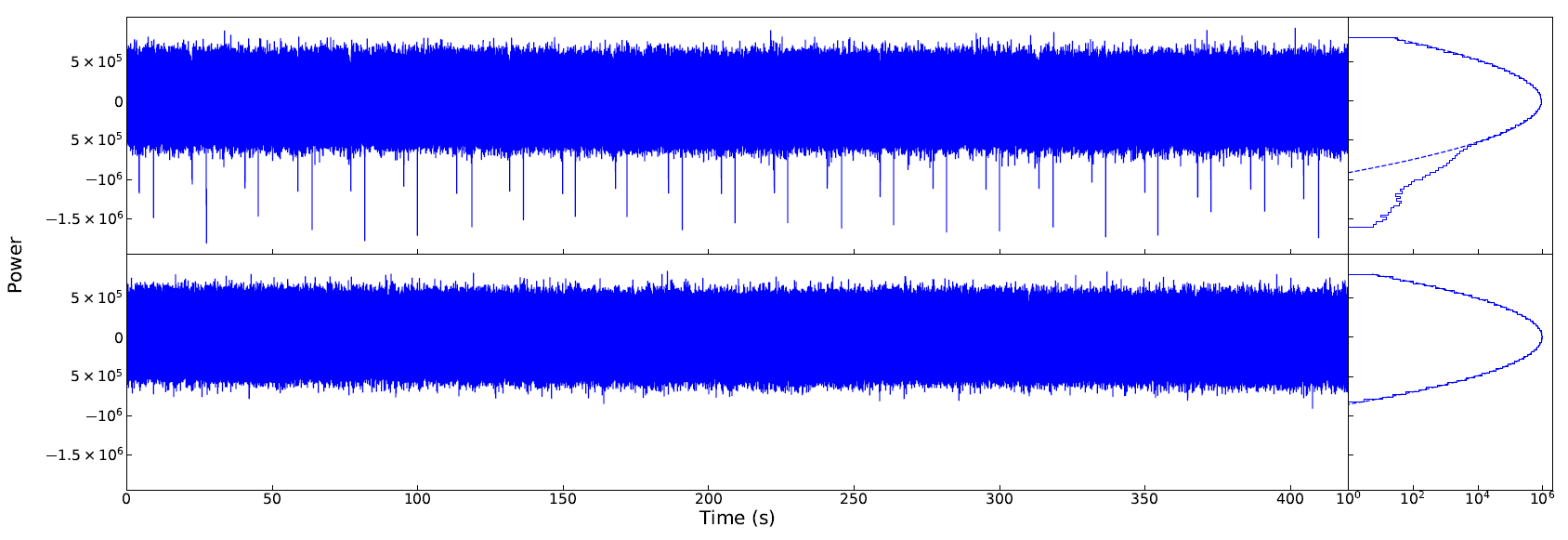}
\caption{Timeseries from one 7-min scan on Sgr~A*, with the mean power subtracted. The upper panel shows the original data where power drop-offs are seen to occur repetitively every 18.192\,s. Histogram of the time samples exhibits a clear deviation at the negative end from a best-fitted Guassian function. The lower panel shows the same chunk of data after application of the cleaning scheme, where the statistics of the time samples is well fitted with a Gaussian distribution. \label{fig:ts-no143}}
\end{figure*}

\begin{figure}
\centering
\includegraphics[scale=0.35]{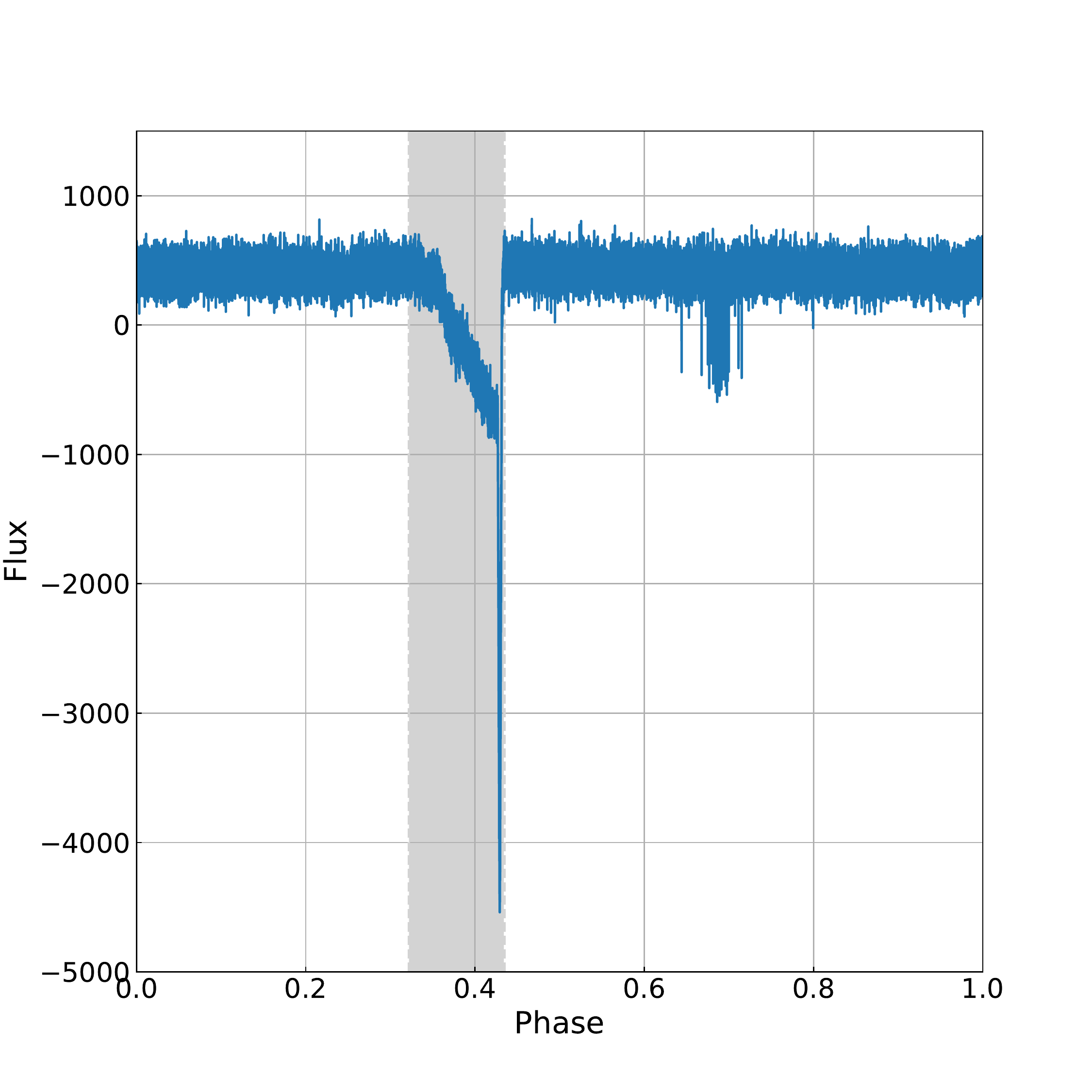}
\caption{Averaged profile of the timeseries in Figure~\ref{fig:ts-no143} with respect to the period of phasing cycle in APP. The power drop-off around phase 0.4 coincides with the dump time in between two sub-scans and its phase range is marked in grey color. The drop-offs around phase 0.7 are associated with the injection of a new phasing solution into the system. \label{fig:dip-prof}}
\end{figure}

The raw baseband voltage data collected during the GMVA observations were converted into intensity timeseries as part of the data processing offline. Figure~\ref{fig:ts-no143} (upper panel) shows the timeseries from one of the scans as an example. A cyclic power drop-off feature is immediately noticeable on top of the overall baseline, and was found to be of the same period as the phasing cycle of the APP which is 18.192\,s. Figure~\ref{fig:dip-prof} shows the feature averaged with respect to its period over the entire time span in Figure~\ref{fig:ts-no143}. It can be seen that in every cycle, the occurrence of power drop-off clusters within two phase ranges (around 0.4 and 0.7) which in turn have been corresponded to a 2.064-s ``dump time'' in between 16.128-s observing sub-scans and the refresh of the phasing solution, respectively \citep[detailed in][]{gmm+19}. A zoomed browse through the timeseries has found that during the dump time, the baseline drops progressively in different number of steps each cycle, but always recovers within a short time window in the end. This is also indicated by the steep edge at the trailing side of the averaged feature in Figure~\ref{fig:dip-prof}. Thus, the trailing edge can be used as the reference to identify the phase range of the dump time (shown as the filled region in Figure~\ref{fig:dip-prof}) which is not known beforehand. The power drop-offs associated with injection of a new phasing solution are in general 2-4\,s trailing the dump time but not precisely periodic since the computing time cost to obtain the solution is different in every cycle. Nonetheless, the drop-off usually exhibits as a simple negative step function without significantly altering the level of baseline afterwards. 

Accordingly, we have established a dedicated scheme to mitigate those systematics in the data. For every scan, we first folded the timeseries with respect to the phasing period and used the averaged feature from dump time to determine its absolute phase. Then we returned to the timeseries and replaced the samples during the dump time with random noise created using mean and rms obtained from the sub-scan right before. The power drop-off caused by the refresh of phase solution can be effectively cleaned with the built-in clip function in \textsc{presto}'s \texttt{prepdata} program while creating the timeseries. An example of the cleaned data after applying the scheme can be found in the lower panel of Figure~\ref{fig:ts-no143}. It is clearly seen that the vast majority of the systematics has been removed, and the distribution of time samples is well modelled by a Gaussian function.

\section{Baseline variation in Stokes} \label{app:sto}
Here we investigate the property of polarization components when the input signal of a data-recording system has a time variability as a consequence of e.g., fluctuation in system temperature. For simplicity, we consider an ideal orthogonal system where the two linear feeds are equally illuminated by the input signal. The signals from the two polarization channels can be expressed as:
\begin{eqnarray}
\displaystyle\left\{\begin{array}{c} x=\hat{a}_{\rm x}+\hat{b}_{\rm x}\textbf{i} \\
\displaystyle y=\hat{a}_{\rm y}+\hat{b}_{\rm y}\textbf{i},
\end{array}\right.
\label{eq:signal}
\end{eqnarray}
where $\hat{a}_{\rm x}$, $\hat{b}_{\rm x}$, $\hat{a}_{\rm y}$, $\hat{b}_{\rm y}$ are all independent variables. Assuming that the input signal is pure Gaussian noise with time-dependent variance, we have:
\begin{eqnarray}
\displaystyle\left\{\begin{array}{l}
\overline{\hat{a}_{\rm x}}=\overline{\hat{b}_{\rm x}}=\overline{\hat{a}_{\rm y}}=\overline{\hat{b}_{\rm y}}=0, \\
\overline{\hat{a}^2_{\rm x}}=\overline{\hat{b}^2_{\rm x}}=\overline{\hat{a}^2_{\rm y}}=\overline{\hat{b}^2_{\rm y}}=\sigma^2(t),
\end{array}\right.
\label{eq:signalstat}
\end{eqnarray}
where $\sigma^2(t)$ is the actual detected power as a function of time. Accordingly, the four Stokes parameters are written as:
\begin{eqnarray}
\displaystyle\left\{\begin{array}{l}
\overline{I}=\overline{xx^*+yy^*}=4\sigma^2(t) \\
\overline{Q}=\overline{xx^*-yy^*}=0 \\
\overline{U}=\overline{2\Re(x^*y)}=\overline{2\hat{a}_{\rm x}\hat{a}_{\rm y}}+\overline{2\hat{b}_{\rm x}\hat{b}_{\rm y}}=0 \\
\overline{V}=\overline{2\Im(x^*y)}=\overline{2\hat{a}_{\rm x}\hat{b}_{\rm y}}+\overline{2\hat{a}_{\rm y}\hat{b}_{\rm x}}=0
\end{array}\right.
\label{eq:signalstat}
\end{eqnarray}
Thus, while the power in $I$ changes as a function of time, there is no resulting time variability in the other Stokes parameters, $Q$, $U$ and $V$, as far as the input signal is unpolarized. Meanwhile, the detection in linear polarization $L$, is written as:
\begin{equation}
\overline{L}=\sqrt{\overline{L^2}}=\sqrt{\overline{Q^2+U^2}}=4\sigma^2(t),
\end{equation}
where
\begin{eqnarray}
\overline{Q^2}=\overline{(xx^*-yy^*)^2}=\overline{\hat{a}^4_{\rm x}}+\overline{\hat{b}^4_{\rm x}}+\overline{\hat{a}^4_{\rm y}}+\overline{\hat{b}^4_{\rm y}}=8\sigma^4(t), \\
\overline{U^2}=\overline{[2\Re(x^*y)]^2}=\overline{\hat{a}^2_{\rm x}\hat{a}^2_{\rm y}}+\overline{\hat{b}^2_{\rm x}\hat{b}^2_{\rm y}}=8\sigma^4(t).
\end{eqnarray}
Therefore, the detection of linear polarization will also be affected when the power in $I$ exhibits a time variability. 

\bibliographystyle{aasjournal}
\bibliography{journals,psrrefs,modrefs,crossrefs}

\end{document}